\documentclass[aps,prx,twocolumn,superscriptaddress,notitlepage,floatfix]{revtex4-2}
\pdfoutput=1
\usepackage[latin9]{inputenc}
\usepackage{tabularx}
\usepackage{blindtext}
\usepackage{ragged2e}
\usepackage{subfig}
\usepackage{color}
\usepackage{amsmath}
\usepackage{caption}
\usepackage{color}
\usepackage{amsthm}
\usepackage{amssymb}
\usepackage{graphicx}
\usepackage{float}
\usepackage{braket}
\usepackage{dsfont}

\usepackage{algorithm}  
\usepackage{algorithmicx}  
\usepackage{algpseudocode}

\usepackage[unicode=true,
 bookmarks=true,bookmarksnumbered=false,bookmarksopen=false,
 breaklinks=false,pdfborder={0 0 1},backref=false,colorlinks=true]
 {hyperref}
\hypersetup{
 linkcolor=magenta, urlcolor=blue, citecolor=blue, pdfstartview={FitH}, hyperfootnotes=false, unicode=true}

\makeatletter
\usepackage{amsfonts,amsthm}\usepackage{tabularx}\usepackage{dcolumn}\usepackage{bm}\usepackage{graphicx}\usepackage{epstopdf}
\usepackage{times}
\hypersetup{urlcolor=blue}

\def\ket#1{\left|#1\right\rangle}

\makeatother

\begin{document}
%\title{Quantum Universal Attack of Heterogeneous Classification Tasks}
\title{Universal adversarial perturbations for multiple classification tasks with quantum  classifiers}

\affiliation{Department of Computer Science and Engineering, South China University of Technology}
\author{Yun-Zhong Qiu\\ \href{mailto:202030442205@mail.scut.edu.cn}{202030442205@mail.scut.edu.cn}}
\begin{abstract}
    Quantum adversarial machine learning is an emerging field that studies the vulnerability of quantum learning systems against adversarial perturbations and develops possible defense strategies. Quantum universal adversarial perturbations are small perturbations, which can make different input samples into adversarial examples that may deceive a given quantum classifier. This is a field that was rarely looked into but worthwhile investigating because universal perturbations might simplify malicious attacks to a large extent, causing unexpected devastation to quantum machine learning models. In this paper, we take a step forward and explore the quantum universal perturbations in the context of heterogeneous classification tasks. In particular, we find that quantum classifiers that achieve almost state-of-the-art accuracy on two different classification tasks can be both conclusively deceived by one carefully-crafted universal perturbation. This result is explicitly demonstrated with well-designed quantum continual learning models with elastic weight consolidation method to avoid catastrophic forgetting, as well as real-life heterogeneous datasets from hand-written digits and medical MRI images. Our results provide a simple and efficient way to generate universal perturbations on heterogeneous classification tasks and thus would provide valuable guidance for future quantum learning technologies. 
\end{abstract}

\maketitle

\section{Introduction}
The interplay between quantum computing and machine learning gives rise to a research frontier of quantum machine learning \cite{sarma2019machine,cerezo2022challenges}.  The attempting quantum computing characteristics hold the intriguing potential to trigger a revolution in traditional machine learning study \cite{li2022recent,huang2022quantum,pan2022deep,wu2023quantum,biamonte2017quantum,xiao2022parameter,huang2022provably,xiao2023practical}. Along this direction, a series of careful investigations have been conducted
and various kinds of quantum classifiers have been introduced in Refs. \cite{oh2020tutorial,li2022quantum,bausch2020recurrent,Liu2022Quantum_}. These results imply potential unparalleled advantages for quantum classifiers compared with their classical counterparts \cite{li2022quantum}.  %In addition, quantum binary classifiers based on the joint framework of Yao \cite{luo2020yao} and Flux \cite{innes2018flux} in Julia programming language \cite{bezanson2012julia} have been 
%Also, in ref.\cite{li2022quantum}, the authors present a well-defined quantum binary classifier based on the joint framework of Yao \cite{luo2020yao} and Flux \cite{innes2018flux} in Julia programming language \cite{bezanson2012julia}. 
The proposed quantum classifiers are typically composed of ``parameterized layers" \cite{li2022quantum} and ``entangled layers" \cite{li2022quantum}. Such classifiers can reach the state-of-art accuracy of over 95\% on binary classification tasks by utilizing limited computational resources in acceptable time complexity. Given the advantages of this design, we implement our classifier with a similar approach.

With the emergence of quantum machine learning methods, a serious concern comes to the reliability of these proposed models. There is sufficient evidence proving that quantum classifiers are as fragile as classical classifiers when encountering adversarial perturbations \cite{Pei-Xin2021Adversarial}. In Refs. \cite{lu2020quantum,liao2021robust,edwards2020quantum}, the authors provide important evidence that quantum classifiers are almost as vulnerable as classical ones. By applying like-wise carefully crafted subtle adversarial perturbations generated by white-box or black-box attack algorithms, the quantum learning systems can be conclusively deceived regardless of the data types (either classical or quantum mechanical). Further studies on the robustness of quantum classifiers have also been carried out in Ref. \cite{guan2021robustness, gong2022enhancing}. Moreover, in Refs. \cite{jiang2019adversarial,zhang2022experimental} a series of specific quantum adversarial machine learning algorithms and physical experiment implementations have been introduced to attack quantum classifiers in different scenarios, including differential evolution algorithm \cite{das2010differential} and zeroth-order optimization \cite{chen2017zoo} for discrete attacks, and fast gradient sign method \cite{goodfellow2014explaining}, projected gradient descent method \cite{madry2017towards}, and momentum iterative method \cite{dong2018boosting} for continuous attacks. 
Such algorithms are well-designed, well-tested, and proven to be robust so that they can generate adversarial samples with high equality. In practice, Ref. \cite{ren2022experimental} performs quantum adversarial learning on programmable superconducting quantum devices, which depict an intriguing blueprint for future quantum devices and applications.

However, most existing works on quantum adversarial learning have focused on adversarial perturbations for single input samples.  In other words, the perturbation for different legitimate images is different, which indicates extra calculation for every image to be perturbed. In classical adversarial machine learning, universal adversarial perturbations have been introduced and investigated in Ref. \cite{moosavi2017universal}. More recently, this has been carried over to the quantum domain in Ref. \cite{gong2022universal}, where the authors studied universal adversarial examples and perturbations. 
This paper not only introduced an effective method for generating universal perturbations for a single classifier but also demonstrated that increasing the strength of perturbations by a certain extent can result in moderate universal adversarial risk. The conclusion in this paper indicates that universal perturbations may exist even for different classification tasks. 
However, a clear demonstration of such universal perturbations remains lacking and there is no known efficient algorithm to find these perturbations hitherto.

In this work, we fill up this gap by studying universal perturbations for multiple classification tasks, in the context of quantum continual learning \cite{jiang2022quantum}. {\color{black} Here, the expression ``multiple classification tasks'' can be more precisely stated as ``heterogeneous classification tasks''. We use the word ``heterogeneous'' to indicate datasets that are generated from different patterns. Since we want to study universal perturbations for multiple classification tasks, the datasets of different tasks should be independently distributed. For example, we take two MNIST handwritten digits classification tasks as ``homogeneous'' because both datasets are generated from the same pattern. Additionally, we take the MNIST handwritten digit and the MedNIST MRI image classification tasks as ``heterogeneous'' because both datasets are generated from different patterns. For the sake of simplicity, we shall use the phrase ``multiple classification tasks'' to represent ``heterogeneous classification tasks''.}

Specifically, our goal is to find an efficient way to generate universal perturbations for quantum classifiers tackling multiple classification tasks.  We simulate two classifiers by merging them using quantum continual learning with elastic weight consolidation (EWC) \cite{mccloskey1989catastrophic,robins1995catastrophic,french1999catastrophic} techniques, then conduct a simple basic iteration method to attack the merged classifier. We find that universal perturbations that can make different input samples into adversarial examples for different classification tasks indeed exist and can be obtained efficiently by the iteration algorithm. In particular, we can obtain a robust multiple classification tasks classifier with a stable average accuracy of over $90\%$ and a loss of around $0.62$. By adding the universal perturbation with strength $0.02$, the average accuracy decreases from $93.3\%$ to $28.5\%$ with fidelity of $0.79$ and loss of around $0.8$. Specifically, the previously trained task's accuracy decreases from $94.5\%$ to $24.5\%$ with fidelity of $0.84$ and loss of $0.73$. The later trained task's accuracy decreases from $92.0\%$ to $32.5\%$ with fidelity of $0.76$ and loss of $0.70$. Our results reveal the universal feature for adversarial perturbations from a new multi-task perspective, which would be useful for future quantum learning technologies.

%The EWC method was originally developed in the context of classical adversarial machine learning to mitigate the problem of catastrophic forgetting \cite{goodfellow2013empirical,kemker2018measuring} and later introduced to quantum adversarial machine learning in Ref.\cite{jiang2022quantum}. For the sake of simplicity, we presume that the two target classifiers work on the same size of qubits and circuit depth, all inputs are normalized and of the same pixel size. Based on our numerical results, it can be inferred that the proposed perturbation method effectively deceives both classification tasks with conclusive outcomes.

\section{Concepts and Notations}

\subsection{The existing quantum adversarial attack methods}

Adversarial machine learning has been extensively discussed in the literature, as evidenced by the work by Huang {\it et al.} \cite{huang2011adversarial}. This field can be approached from two distinct perspectives as a ``fact" and an ``optimization problem." The ``fact" perspective highlights the existence of subtle perturbations that, when applied to the input of machine learning models, can significantly modify their outputs. On the other hand, the ``optimization problem" perspective involves transforming the task into a ``maximizing-minimizing problem" and leveraging algorithms with the assistance of computational power to generate adversarial samples. Various logical methods can be employed to attack machine learning systems, resulting in different strategies for adversarial attacks. In this paper, we follow the classification and analysis of adversarial attacks presented in the work by Vorobeychik {\it et al.} \cite{vorobeychik2018adversarial} with a particular focus on supervised learning scenarios.

In the context of supervised learning scenarios, our analysis considers a dataset $D_N$ comprising  $N$ elements, where each element is denoted by $x^{(i)}$ and $y^{(i)}$, representing the input data to be classified and its corresponding label, respectively \cite{lecun2015deep,cunningham2008supervised,osisanwo2017supervised}. To ensure compatibility with the quantum classifier, the input data $x^{(i)}$ needs to be transformed into a valid quantum state $\ket{x^{(i)}}$ before the classification process. Assuming that we have trained a quantum classification model $F$ to achieve high accuracy on dataset $D_N$, we can affirm that for every input data element $x^{(i)}$, the discrepancy between the model's output $F\left(\ket{x^{(i)}}\right)$ and ground truth label $y^{(i)}$, which is calculated from the loss function $L$, is constrained within an acceptable range $\epsilon$: 

\begin{equation}
    \sum_{i=1}^n L\left(F\left(\ket{x^{(i)}}\right),y^{(i)}\right)\le \epsilon.
\end{equation}

From an alternative standpoint, the training process in machine learning predominantly focuses on minimizing the loss function. In contrast, the adversarial process aims to maximize the loss function by applying a perturbation $\delta^{(i)}$ within the constraints of a small region $\Delta$. A commonly employed technique involves imposing an $l_p$-norm bound, which restricts the magnitude of the perturbation. Consequently, the aforementioned description can be summarized as follows:
\begin{equation}
    \begin{aligned}
    &\max_{\delta^{(i)}}\quad L\left(F\left(\ket{x^{(i)} + \delta^{(i)}}\right) , y^{(i)}\right) \\
    &\text{s.t.}\quad  \left(\sum_{i=1}^{N}||\delta^{(i)}||^p\right)^{\frac{1}{p}} \le\Delta    
    \end{aligned}.
\end{equation}
Intuitively, the objective of this optimization problem is to achieve an imperceptible perturbation that generates a significant change in the loss function. An intuitive approach is to apply the perturbation along the maximum gradient ascent direction, as the gradient of the loss function indicates the rate of change in a given direction. Suppose we define the function $J(x^{(i)})$ that calculates the absolute gradient value and ascending direction of the gradient for $x^{(i)}$ in a multi-dimensional Hilbert space, as well as the strength of the perturbation denoted by $\epsilon$. In this context, the calculation of perturbation $\delta^{(i)}$ can be summarized as follows:
\begin{equation}
    \delta^{(i)} = \epsilon\times \mathrm{sgn}(J(x^{(i)})).
\end{equation}

Most white-box attacks, such as FGSM \cite{goodfellow2014explaining} (fast gradient sign method), PGD \cite{madry2017towards} (projected gradient descent), and MIM \cite{dong2018boosting} (moment iteration method), draw inspiration from directly calculating gradients and selecting an appropriate perturbation. Meanwhile, for black-box attacks, direct access to the gradient is not available. Substitute optimization methods like differential evolution algorithms are employed to identify the pixels that can significantly alter the value of the loss function. In quantum adversarial machine learning, gradient plays a crucial role as it indicates the regions where the loss function can be easily manipulated. The utilization of gradient information enables us to create minor perturbations specifically in regions characterized by higher absolute gradient values. These perturbations cause a shift in the loss function, guiding the model toward a desired direction and leading to a significant alteration in the model's performance, thereby producing an adversarial effect.

\subsection{The universal adversarial perturbation}
According to the aforementioned description, the perturbation $\delta^{(i)}$ is calculated individually for each input sample $x^{(i)}$. However, this approach requires attackers to compute specific information for each input, which can be challenging in many attacking scenarios. To address this limitation, a new type of adversarial method called ``universal perturbation" has been developed. The underlying principle is to exploit the uniformity within a single task, identify similarities in gradient information across selected samples, and apply the same universal perturbation to transform most original samples into adversarial samples. In other words, the objective is to find a universal perturbation $\delta$ that combines the perturbation strength $\epsilon$ and the gradient sign information $J$ from the entire selected dataset $X$:
\begin{equation}
    \delta = \epsilon\times \mathrm{sgn}(J(X)).
\end{equation}
By adding this universal perturbation to the original image, we expect to achieve the adversarial transformation of multiple input samples simultaneously.

Extensive evidence has convincingly shown the effectiveness and potential risks associated with universal perturbations for quantum classifiers \cite{gong2022universal}. Taking this understanding a step further, we pose an intriguing question: Can we generate a universal perturbation capable of deceiving a set of classifiers performing different tasks? To simplify the scenario, consider two classifiers, $A$ and $B$. Their respective datasets are denoted as $X_A$ and $X_B$. Additionally, we define the function $J$ as the operator that captures gradient information. The formulation of the universal perturbation can be expressed as follows:

\begin{equation}
    \delta = \epsilon\times \mathrm{sgn}(J(X_A,X_B)).
\end{equation}

In scenarios where gradient information is not exclusively associated with a single input sample, it becomes unavoidable that the resulting perturbation encompasses distributed information from the entire selected dataset, which may not perfectly align with each individual input sample. To overcome this challenge, it is essential to employ larger perturbation strengths and enhance the functionality of gradient information-extraction methods. By doing so, we can effectively address the issue of incorporating dataset-wide information and improve the alignment between perturbations and individual input samples.

In order to differentiate between adversarial examples generated using gradient information from individual input samples and those created using universal perturbations, we propose the term ``universal examples" to refer specifically to adversarial examples crafted using universal perturbations. Consequently, we define the adversarial attack that utilizes universal examples as a ``universal attack". This terminology serves the purpose of distinguishing between various approaches and strategies employed in the field of adversarial machine learning. By introducing these specific terms, we can more precisely categorize and discuss the different techniques and methodologies involved in adversarial machine learning.

\subsection{Quantum continual learning and catastrophic forgetting}
The design of algorithms to implement the concept of finding universal perturbation poses a significant challenge. Currently, performing calculations on multiple quantum classifiers with different structures is difficult due to limitations in qubit volume and incoherences arising from inconsistencies of quantum circuit parameters. To address this issue, we propose to introduce the quantum continual learning method to merge two target classifiers into one robust continual learning classifier that is capable of tackling multiple classification tasks. Provided that a universal perturbation can be found on this continual learning classifier, the argument of effective universal perturbation of multiple classification tasks shall still hold. Intuitively, quantum continual learning techniques offer a powerful approach to simplify the generating process of universal perturbation and demonstration of universal vulnerability between different classification tasks.

Quantum continual learning, as explored in the work by Jiang {\it et al.} \cite{jiang2022quantum}, encompasses the study of training a classifier capable of performing multiple tasks using the same set of parameters. However, a significant issue arises when training a classifier on two datasets sequentially, as it often results in a substantial decrease in accuracy for the previously trained task. This phenomenon is known as ``catastrophic forgetting" \cite{dunjko2018machine}, which occurs due to the propensity of the learning process for the second task to modify essential parameters associated with the first task. Consequently, the acquisition of robust quantum continual learning models becomes crucial in the context of generating universal perturbations. 

Recent research has indicated that animals encounter similar challenges of forgetting when acquiring new knowledge. Interestingly, a method that involves safeguarding specific excitatory synapses that play a crucial role in preserving past experiences has been discovered \cite{yang2009stably}. To address the issue of catastrophic forgetting, a straightforward approach is to protect the crucial parameters associated with the first task while training the second task. This concept has inspired the development of an effective method in both classical and quantum classifiers to mitigate catastrophic forgetting, known as the elastic weight consolidation (EWC) method \cite{kirkpatrick2017overcoming}. Consider two independent classification tasks, denoted as $A$ and $B$. The training process can be viewed as maximizing the likelihood function $P(\theta | X)$, where $\theta$ represents the parameter being trained. By employing simple Bayes rules \cite{stone2013bayes}, a transformation can be applied as follows:

\begin{equation}
    \mathrm{log}P(\theta|X) = \underset{\text{element 1}} {\mathrm{log}P(X_B|\theta)} +\underset{\text{element 2}}{\mathrm{log}P(\theta|X_A)} - \underset{\text{element 3}}{\mathrm{log}P(X_B)},
\end{equation}
where the first element corresponds to the loss function $L_B(\theta)$ associated with task B, while the third element remains constant with respect to $\theta$. Consequently, the problem can be reformulated as an optimization problem involving only the first two elements. From a technical standpoint, calculating the posterior probability $\log P(\theta | X_A)$ directly can be challenging. To simplify the calculation, the prior probability can be simplified by Gaussian distribution, as suggested in previous studies in Ref. \cite{chopin2011fast}. In addition, we utilize a second-order Taylor expansion, as described by Pourahmadi Ref. \cite{pourahmadi1984taylor}, centered around the optimal parameter $\theta_A$ obtained from the previously trained task $A$. This expansion allows us to approximate the loss function of the current task using the Hessian matrix $H_{\theta_A}$ and neglect higher-order terms, as proposed in \cite{jiang2022quantum}: 
\begin{equation}
    \mathrm{log}P(\theta|X_A)=\mathrm{log}P(\theta_A|X_A)+\frac{1}{2}(\theta-\theta_A)^T H_{\theta_A} (\theta-\theta_A).
\end{equation}
By incorporating the Hessian matrix, we can effectively capture the curvature of the loss landscape and make more precise adjustments to the model's parameters during the continual learning process. This approach helps mitigate the interference caused by the introduction of new tasks, allowing for smoother transitions and improved preservation of previously learned knowledge. This approximation allows us to simplify the calculations and make the optimization process more manageable in practice.

The Hessian matrix \cite{thacker1989role,mizutani2008tutorial} is a square matrix composed of second-order derivatives of a scalar-valued function. It provides valuable information about the local curvature of the function when it has multiple variables. Furthermore, it is important to highlight that the expectation value of the Hessian matrix is equivalent to the Fisher information matrix, as elucidated in the tutorial by Ly {\it et al.} \cite{ly2017tutorial}. The Fisher information matrix plays a fundamental role in statistical learning and is widely employed in various statistical estimation and inference tasks \cite{ly2017tutorial,kunstner2019limitations,frieden2000physics}. Recently, these methods have been extended to the quantum domain \cite{petz2011introduction,liu2020quantum}. Here we introduce the Fisher information matrix to this approximation for three major reasons mentioned in \cite{kirkpatrick2017overcoming}. Firstly, it provides insights into the second derivative of the loss function around a minimum point, offering information about the local curvature. Secondly, it can be computed solely from first-order derivatives, making it computationally feasible for large models. Additionally, the Hessian matrix is guaranteed to be positive and semi-definite. Leveraging these characteristics, we can express the loss function of task $B$ by incorporating a diagonal precision matrix, which is constructed using the diagonal elements $F_i$ of the Fisher information matrix $F$. To control the impact of the EWC constraint, we introduce the hyper-parameter $\lambda$ to represent its strength:
\begin{equation}
    L(\theta) = L_B(\theta)+\frac{\lambda}{2}\sum_iF_i(\theta_i-\theta_{A,i})^2.
\end{equation}

The primary objective of quantum continual learning is to achieve satisfactory performance on task $B$ while preserving relatively high performance on the original task $A$. To accomplish this goal, the EWC method introduces regularization to the original loss function $L(B)$ to penalize deviations from the optimal solution obtained for task $A$, taking into account the varying importance of different parameters. Specifically, the importance of the quantum classifier parameters is assessed using the Fisher information matrix. The diagonal elements of this matrix serve as weights for the penalty term. Intuitively, each Fisher information matrix corresponds to a Hessian matrix, with the diagonal elements representing the local curvature landscape along different variables and directions. These curvature landscapes are closely connected to the concept of gradients: larger curvatures indicate greater significance, suggesting that even slight shifts in a small range can result in significant differences in the value of the loss function.

In other words, parameters with larger curvatures should be less susceptible to modifications during the continual learning process. By incorporating the penalty based on the Fisher information matrix, the EWC method effectively protects these important parameters from catastrophic forgetting, allowing the model to retain valuable knowledge from task $A$ while adapting to new tasks. This explanation provides a high-level overview of the concepts behind quantum continual learning and how the EWC method mitigates the issue of catastrophic forgetting.

\section{Method for finding quantum universal perturbation}

In our study, we aim to leverage the methods discussed in the previous section to generate universal perturbations that can deceive a set of classifiers performing different classification tasks. Typically, universal perturbations and attacks require specific sets of qubits and circuit parameters as contextual information. To simplify the analysis and strengthen the validation of the universal vulnerability of quantum classifiers, we focus on classifiers with the same number of qubits and identical circuit parameter structures, which we refer to as ``homogeneous classifiers."

By ensuring that the classifiers have the same number of qubits and circuit parameters, we ensure that they operate within the same Hilbert space and possess comparable learning capabilities. This allows us to suppress the interference caused by variations in qubit size and circuit structure. By working with ``homogeneous classifiers'', we can confidently evaluate and demonstrate the universal vulnerability of quantum classifiers, as any observed universal perturbation or attack can be attributed to the shared characteristics of the classifiers rather than differences in their underlying architectures.

The methodology we employ is straightforward and allows us to approximate the effect of attacking two ``homogeneous classifiers" as attacking a single classifier capable of performing two classification tasks. To demonstrate the universality of the vulnerability of quantum classifiers, we perform quantum adversarial machine learning algorithms on a quantum continual learning classifier that is capable of handling both tasks.

We start by training a robust quantum classifier using dataset $X_A$ for the first classification task, denoted as task $A$. This initial training process aims to establish a solid foundation for the classifier's performance on task $A$. Subsequently, we apply the method of quantum continual training, along with the EWC method, to mitigate catastrophic forgetting. This approach involves training a new classifier while protecting the important parameters learned from task $A$, denoted as $\theta_A$. The process of continual learning can be intuitively understood as a merging process of the two classifiers, where the knowledge from task $A$ is retained while incorporating new knowledge from task $B$. By the end of the continual learning process, we should obtain a mixed model with parameters $\theta_{AB}$ that can effectively classify both datasets $X_A$ and $X_B$ with high accuracy, typically around 90\%.

Finally, to generate the universal perturbation, we utilize the ``quantum-adapted basic iterative method (qBIM)" in conjunction with the average gradient computed from the entire dataset. The qBIM algorithm leverages the gradient information to iteratively generate perturbations that are universally effective across the trained classifiers, thereby showcasing the vulnerability of the quantum classifiers. By following this methodology, we can demonstrate the generation of universal perturbations and validate the vulnerability of quantum classifiers in the context of ``homogeneous classifiers" performing different classification tasks.

\subsection{Training a quantum continual learning model}
Based on the theoretical analysis of continual learning, the classification tasks of the two classifiers must be independently distributed. To demonstrate the effectiveness of our approach, we have selected two distinct datasets: MNIST handwritten digits and MedNIST medical images.

The MNIST dataset, introduced by LeCun {\it et al.} \cite{lecun2010mnist}, consists of grayscale images of handwritten digits ranging from $0$ to $9$. On the other hand, the MedNIST dataset \cite{medmnistv1,medmnistv2} contains medical images of different body parts, such as the hand and breast. These two datasets are chosen due to their minimal correlations, making them suitable for approximating independent distributions. To ensure fairness in our experiments, we assume that the collected dataset samples from both MNIST and MedNIST are of the same size. We specifically select images of digits $1$ and $9$ from the MNIST dataset and corresponding images of the hand and breast from the MedNIST dataset. This enables us to create a binary classification task for our quantum classifier.

Before feeding the images into a quantum circuit, we preprocess them by rescaling their size from $28\times28$ to $64\times64$. This ensures that we make full use of the available 12 active qubits and enables the images to be represented as logical wave functions, which can be input into the quantum circuit. Additionally, we normalize the images to ensure consistent and effective processing. Furthermore, to facilitate the calculation of cross-entropy loss, we convert the labels into a one-hot encoding format. In total, we create a dataset sample consisting of 1000 training samples and $200$ testing samples for constructing our classifier. By utilizing these datasets and following the aforementioned preprocessing steps, we can proceed with training and evaluating the performance of our quantum classifier on the binary classification task involving MNIST and MedNIST images.

To construct our quantum classifier, we utilize the Yao framework \cite{luo2020yao}, which provides a platform for building quantum circuits. Our classifier consists of both parameterized layers and entangled layers. The parameterized layers in our quantum classifier are composed of the Rz-Rx-Rz circuit applied to each active qubit in the circuit. This circuit structure allows us to introduce rotations along the X-axis and Z-axis of the quantum state's Bloch sphere. These rotations play a crucial role in manipulating the quantum state and extracting relevant information for classification. On the other hand, the entangled layers in our classifier involve the application of CNOT gates. These gates create entanglement between every qubit in the circuit, enabling the qubits to share and distribute quantum information across the system. This entanglement is instrumental in capturing complex relationships and correlations among the qubits, enhancing the expressive power of our quantum classifier \cite{PhysRevX.7.021021}. For our specific implementation, we choose a quantum classifier with a depth of $20$, utilizing $12$ qubits. Each depth corresponds to one parameterized layer followed by one entangled layer. By increasing the depth, we enable the classifier to capture more intricate patterns and dependencies within the input data. The chosen structure, proven to be robust and efficient in quantum machine learning, ensures that our quantum classifier is capable of effectively handling the classification tasks at hand. 

The training process of our quantum classifier involves several steps, starting with training on the MNIST dataset and then applying the continual learning method to train on the MedNIST dataset. Throughout the process, we perform measurements on the last qubit of the quantum circuit, specifically qubit number $12$, using single qubit state operators in the form of one-hot diagonal elements. Each operator represents a specific class in the classification task. During the training for a single classification task, we follow a specific set of operations. Firstly, we initialize the parameters of the quantum circuit with random values. Secondly, we record the data from both the MNIST and MedNIST databases as classical complex matrices in ``.mat" files. Before applying the data to the quantum circuit in the Yao platform, we convert it into the form of a quantum state. Thirdly, for each iteration, the circuit calculates the current loss, accuracy, and fidelity value, and computes the gradients using the auto-differentiation functionality. Fourthly, the gradients are applied to update all parameters of the quantum circuit using the Adam optimizer, implemented in Flux.jl. Finally, when the predefined number of iterations is reached, the learning process concludes, returning the trained circuit parameters along with historical accuracy, loss, and fidelity.

In the first learning process on the MNIST dataset, we employ the ``Flux.Adam" optimizer \cite{innes2018flux, kingma2014adam} with a learning rate of $0.005$. The training is performed with a batch size of 100 and for a total of 30 epochs. As a result, we obtain a model with an accuracy of 97.6\% on the MNIST dataset.

In the subsequent classification task on the MedNIST dataset, we employ the EWC method to mitigate catastrophic forgetting. To implement this method, we first need to calculate the Fisher information matrix before the training process. The calculation of the Fisher information matrix can be performed through the following steps:
\begin{enumerate}
    \item Initialize the Fisher matrix as a square matrix of zeros with a size equal to the length of the circuit parameters.
    \item For each input sample in the batch, compute the expectation value, denoted as $\texttt{expect}$, and the gradient, denoted as \texttt{grad}. Then calculate the gradient for the Fisher matrix as $\texttt{grad\_fisher} =  \texttt{grad}/\texttt{expect}$.
    \item Add the value $\texttt{grad\_fisher} \times \texttt{grad\_fisher}^\text{T}$ to the Fisher matrix.
    \item Finally, take the mean value of the Fisher matrix over the batch size to obtain the resulting Fisher information matrix.
\end{enumerate}

The training process for the MedNIST classification task follows a similar approach to the training on the MNIST dataset. However, there are a few differences. Firstly, the initial circuit parameters should be set as the circuit parameters obtained from the previously trained task. Secondly, during training, an additional term called the EWC punishment term, denoted as $\texttt{EWCpunish}$, is introduced. This term incorporates the EWC strength hyperparameter, $\lambda$, the diagonal elements of the Fisher information matrix, $\texttt{fim}$, the current parameters of the continual training circuit, $\texttt{params}$, and the previously trained circuit parameters, $\texttt{pre\_params}$. The expression for $\texttt{EWCpunish}$ is as follows:

\begin{equation}
    \texttt{EWCpunish}=\lambda\times \texttt{fim}\times(\texttt{params}-\texttt{pre\_params}).
\end{equation}

\begin{figure}[tbp]
	\center
        \captionsetup{labelfont=large,textfont=large}
	\subfloat[The training process for the first task]{
        \begin{minipage}[t]{0.48\textwidth}
        \centering  
        \includegraphics[width=\linewidth]{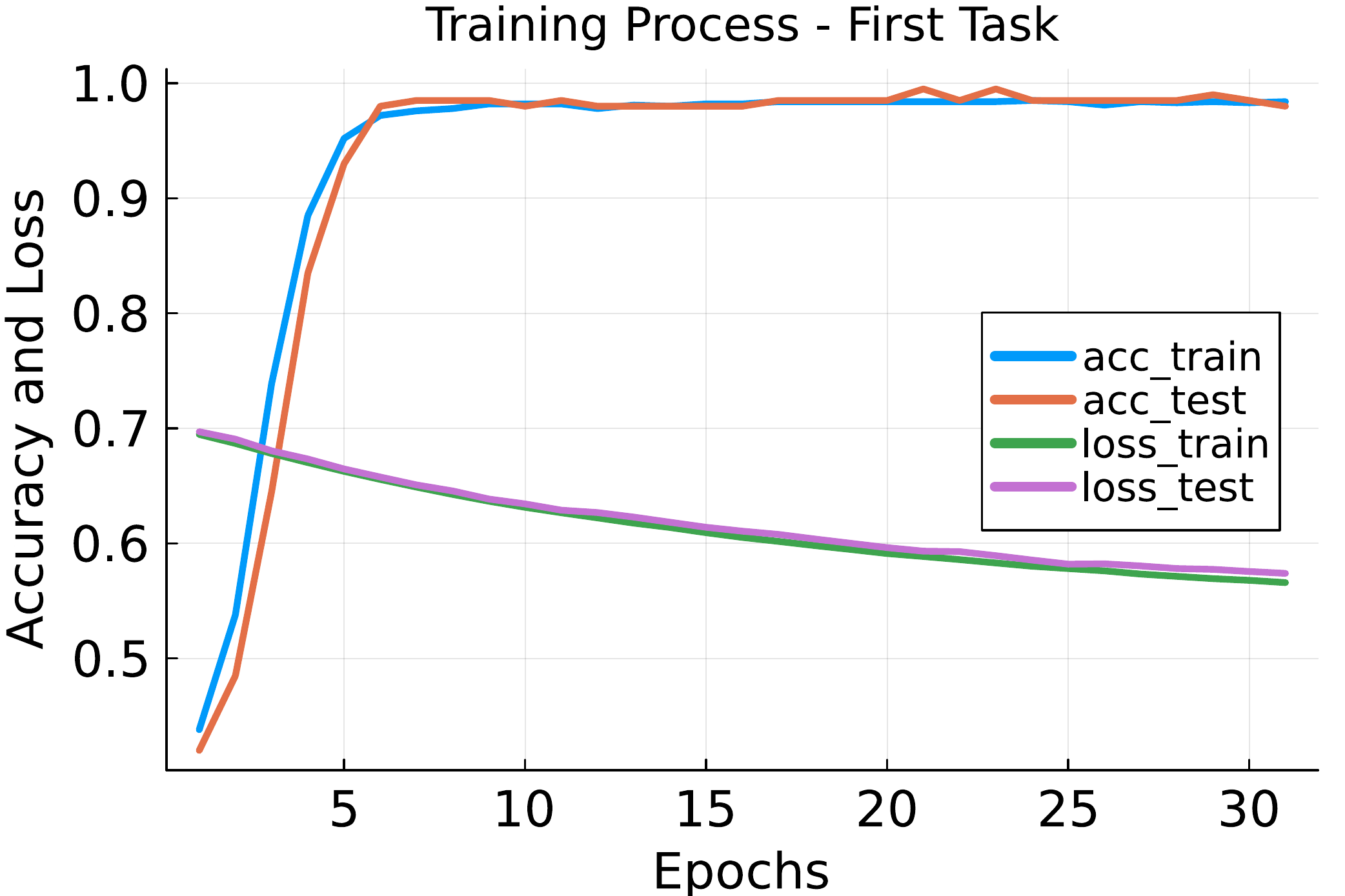}
        \end{minipage}
    }
    \qquad
    \subfloat[The training process for the second task]{
        \begin{minipage}[t]{0.48\textwidth}
        \centering  
        \includegraphics[width=\linewidth]{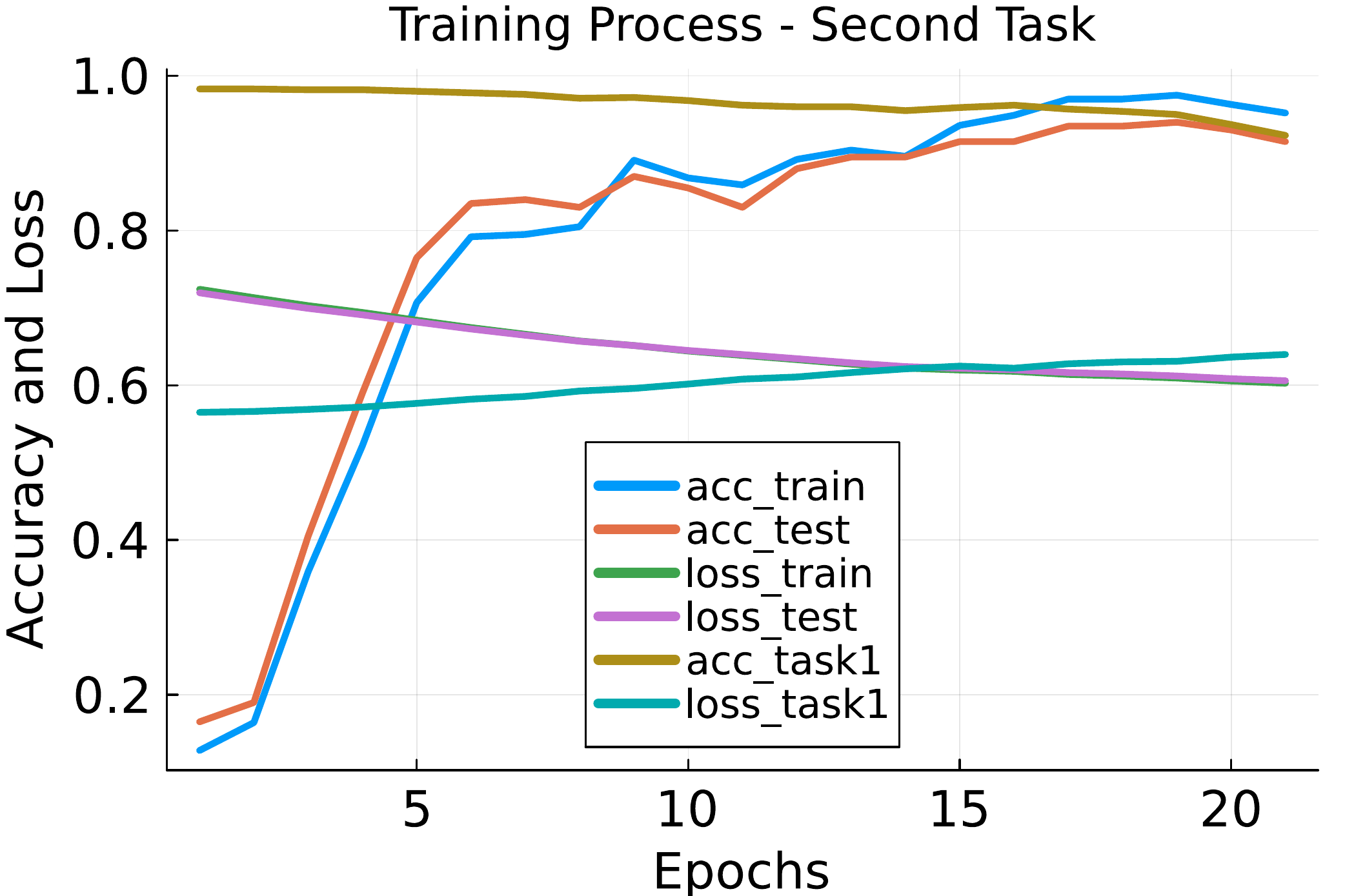}
        \end{minipage}
    }
\captionsetup{justification=RaggedRight,singlelinecheck=false,textfont=normalsize}
\vspace{-10pt}
    \caption{\textbf{The training processes of quantum continual learning.} (a) The first task is the classification of MNIST hand-written digits which end up with an accuracy of around $95\%$. (b) The second task is the classification of MedNIST MRI image training based on the optimized parameters of the first task. By utilizing the EWC method, we acquire an average accuracy of around $90\%$.}
 \label{fig:egfig1}
\end{figure}

In the second learning process, we keep the same learning rate and reduce the training epochs to $20$ to prevent overfitting. During this process, we also apply the EWC method, which introduces an additional parameter, $\lambda$, to achieve the desired effect of continual learning. The choice of the optimal $\lambda$ value is heavily influenced by the characteristics of the dataset and the classifier. In our experiment, we set $\lambda$ to $750$, which resulted in an average accuracy of $93.3\%$ for both classifications. Specifically, the accuracy for the first classification task is $94.5\%$, while the accuracy for the second classification task is $92.0\%$. Despite a slight decrease in the performance of the merged classifier on the first task, both classification accuracies remain at an acceptable level of around $90\%$. The training process is presented in Fig. \ref{fig:egfig1}.

\subsection{Applying quantum Universal Perturbation}

During the perturbation process, we employ the ``quantum basic iteration method (qBIM)" to generate adversarial examples and evaluate the effectiveness of the attack at different perturbation strengths. The qBIM method is a variation of the well-known Fast Gradient Sign Method (FGSM). Unlike FGSM, which generates adversarial examples in a single step, qBIM iteratively applies FGSM with a small step size. This iterative approach allows for adjustments in the perturbation direction during adversarial machine learning, resulting in more robustly generated samples. Let $\epsilon$ denote the step size, $x_k$ represent an arbitrary sample from the collective dataset ${X_A, X_B}$, $\pi_C$ denote the projection operator that normalizes the wave function, and $L$ denote the loss function. At each iteration step $k+1$, the qBIM algorithm can be described as follows:

\begin{equation}
    x_{k+1}=\pi_C[x_k+\epsilon\times\mathrm{sgn}(\nabla L(F(X_A,X_B)))].
\end{equation}

To provide a more detailed explanation, here is a pseudo-code representation of the quantum-adapted basic iterative method (qBIM) for finding effective universal perturbation, as presented in Algorithm \ref{algoBIM}:

\begin{figure}
\begin{algorithm}[H]  %其中这里面不能有H不然会报错，不过不影响结果
\caption{Quantum-adapted basic iterative method}%算法名字
\label{algoBIM}
  \begin{algorithmic}[1]  
	\Require {The trained model $F$,loss function $L$, step size $\epsilon$, number of iteration $n$, legitimate datasets $X_A$ and $X_B$, size of dataset $num$}%输入参数
	\Ensure {Adversarial sample $X_{adv}$}
	\State put the two datasets together into new dataset $X_{all}=(X_A,X_B)$ \;
        \State put $X_{all}$ into quantum state $\ket{X_{all}}$\;
        \State copy variable $X_{all}$ into newly built variable $X_{adv}$\;
        \State put $X_{adv}$ into quantum state $\ket{X_{adv}}$\;
	\For{every sample in collective dataset $X_{all}$}
            \State calculate the gradient $J_i=\nabla L(F(X_A,X_B))$ \;
       \EndFor
       \State gradient information $J=\textbf{mean}(\sum_{i=1}^{num}(J_i))$\;
        \For{$k=1,...,n$}
            \State put $\ket{X_{adv}}$ into classifier to get result $F(\ket{X_{adv}})$\;
            \State calculate the accuracy of $F(\ket{X_{adv}})$ and record\;
            \State calculate the loss of $F(\ket{X_{adv}})$ and record\;
            \State calculate the fidelity between $X_{all}$ and $X_{adv}$ and record\;
            \For{every sample $i$ in collective dataset $X_{all}$}
                \State $X_{adv}[i]=X_{adv}[i]+\epsilon\times\mathrm{sgn}(J)$\;
                \State $X_{adv}[i]/=\mathrm{norms}(X_{adv}[i])$\;
            \EndFor
        \EndFor
	\State return $X_{adv}$\;
    \end{algorithmic}
\end{algorithm}
\end{figure}

We conducted a perturbation experiment on the concatenated dataset of the two trained tasks using $30$ iterations and a total perturbation strength of $0.02$. In each iteration, we calculated the gradient information of the initial unperturbed $1024$ circuit input pixels to determine the perturbation direction. Specifically, we computed the mean value of the pixel-wise gradient for all $400$ samples ($200$ samples for each classification task) to indicate the direction of gradient ascent, which served as the perturbation direction. The perturbation itself was obtained by multiplying the perturbation-per-step value with the sign of the gradient value. 
{\color{black} During each attacking step, we first apply the calculated perturbation to every single sample from two classification tasks to generate universal adversarial examples. Then, we clip the pixel value of universal adversarial examples within the range of $[0,1)$ to avoid meaningless values. Furthermore, we normalize the adversarial example after clipping to ensure that the generated adversarial example is in a legal quantum state. }

%This universal perturbation was then applied to every sample in the concatenated dataset. To ensure the perturbations remained within reasonable bounds and did not lead to unexpected or invalid results, we performed clipping of pixel values within the range of $[0, 1)$ and normalized the perturbed samples to ensure they remained in a valid logical quantum state.

\begin{figure}
   \centering
   \includegraphics[width=\linewidth]{
        {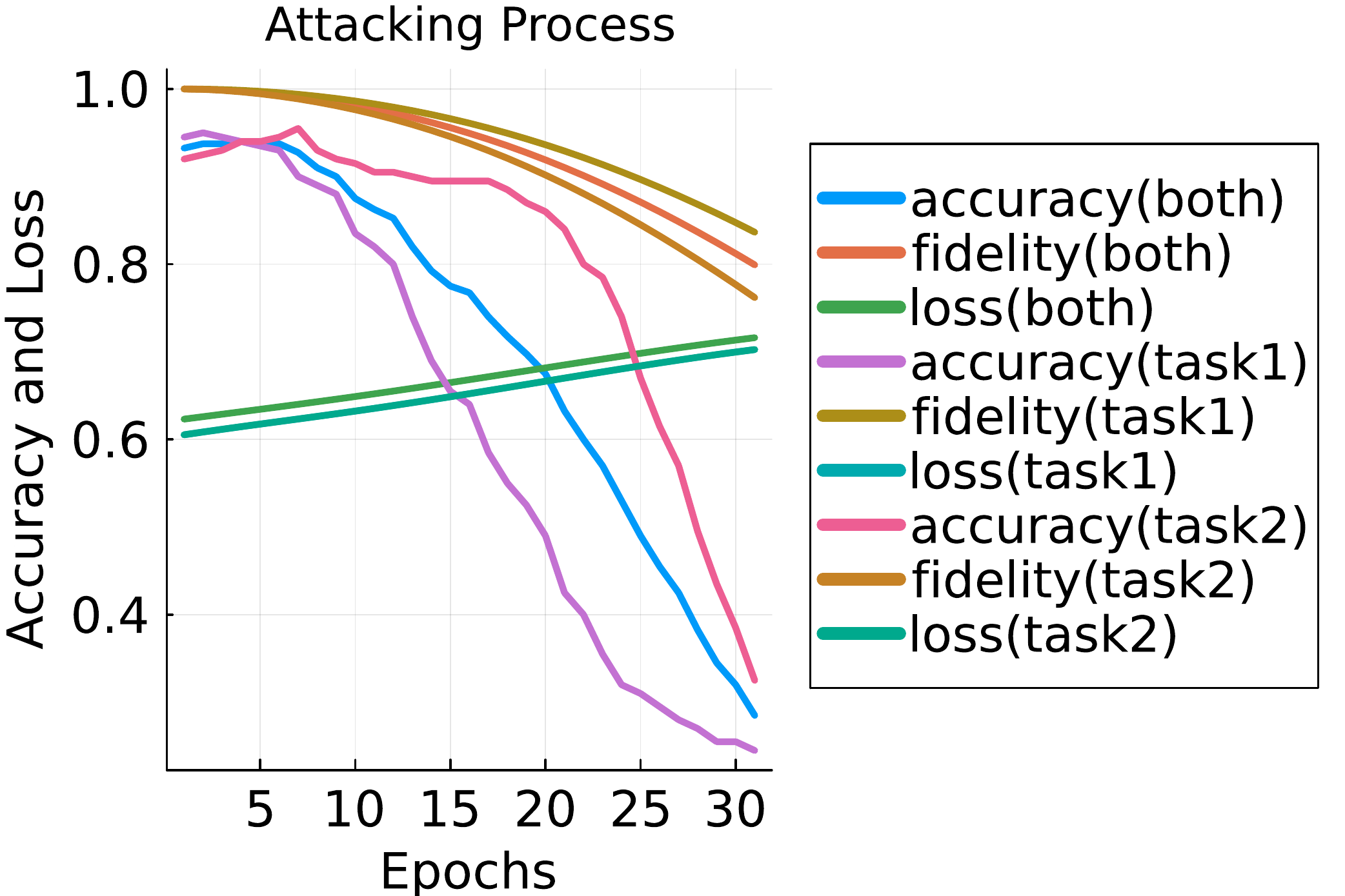}
    }
    \hyphenpenalty=-1
\tolerance=1000
\captionsetup{justification=RaggedRight,singlelinecheck=false,textfont=normalsize,width=\linewidth}
    \caption{\textbf{Quantum universal perturbation} This figure demonstrates the perturbing process of finding quantum universal perturbation on a classifier trained on MNIST and MedNIST datasets using quantum continual learning and EWC techniques. The average and separate accuracy, fidelity, and loss values are presented explicitly. There is not much difference in fidelity and loss, but a prominent distinction in accuracy value between different tasks.}
    \label{fig:egfig2}
\end{figure}

During the perturbation process, the average accuracy of the two datasets decreased from $93.3\%$ to $28.5\%$ with fidelity of $0.79$. Here we present a curve illustration in Fig. \ref{fig:egfig2}. Specifically, the accuracy of the individual tasks dropped from $94.5\%$ and $92.0\%$ to $24.5\%$ (fidelity = $0.84$) and $32.5\%$ (fidelity =$ 0.76$), respectively. This significant reduction in accuracy demonstrates that the universal perturbation generated by the mean value of the gradient for all dataset samples is capable of deceiving the classifier on both classification tasks almost completely. Additionally, it is worth noting that the accuracy and fidelity of the tasks do not change synchronously. The accuracy of the previously trained task experiences a rapid decline at the beginning of the perturbation process and then stabilizes at a higher fidelity. On the other hand, the later-trained task remains less sensitive to perturbation, maintaining a higher accuracy and lower fidelity compared to the previously trained task. This observation suggests that the previously trained task is more vulnerable, as the classifier's performance deteriorates significantly even with a smaller amount of perturbation.

\begin{figure}
\centering
\captionsetup{labelfont=large,textfont=large}
\subfloat[digit 1]{
    \begin{minipage}[h]{0.49\linewidth}
    \centering
    \includegraphics[width=1\linewidth]{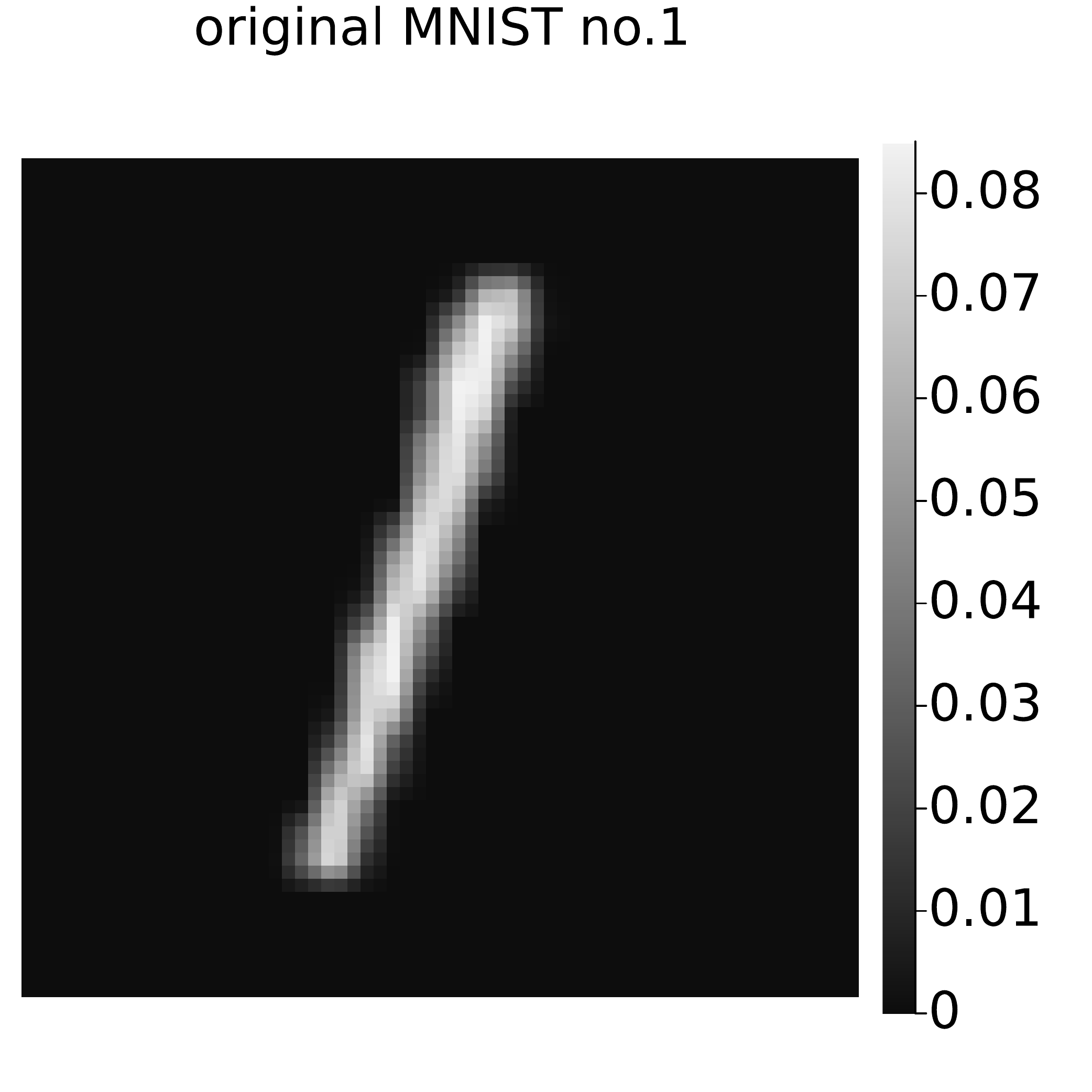}
    \end{minipage}
}
\subfloat[adversarial digit 1]{    
    \begin{minipage}[h]{0.49\linewidth}
    \centering
    \includegraphics[width=1\linewidth]{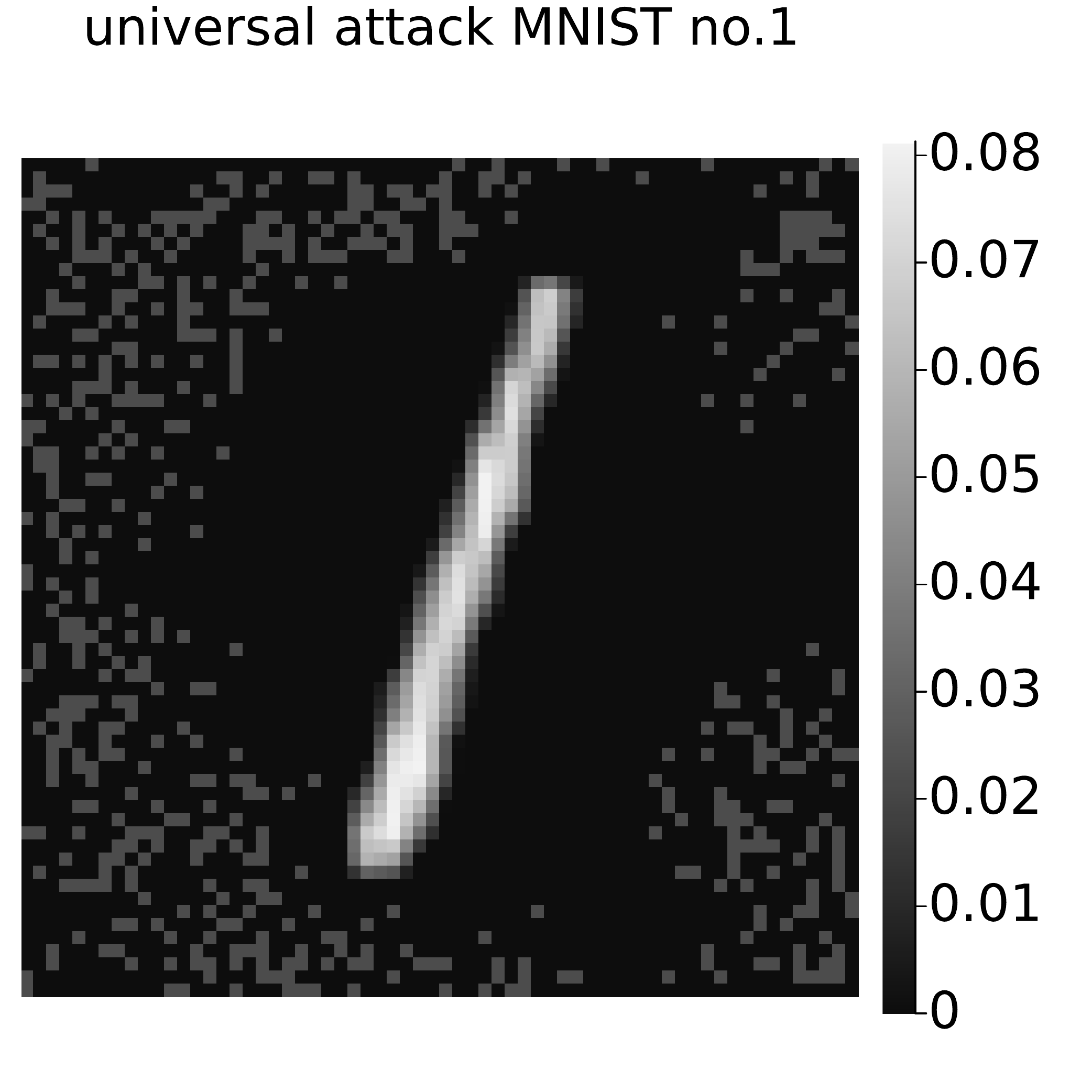}
    \end{minipage}

}
\qquad
\subfloat[digit 9]{
    \begin{minipage}[h]{0.49\linewidth}
    \centering
    \includegraphics[width=1\linewidth]{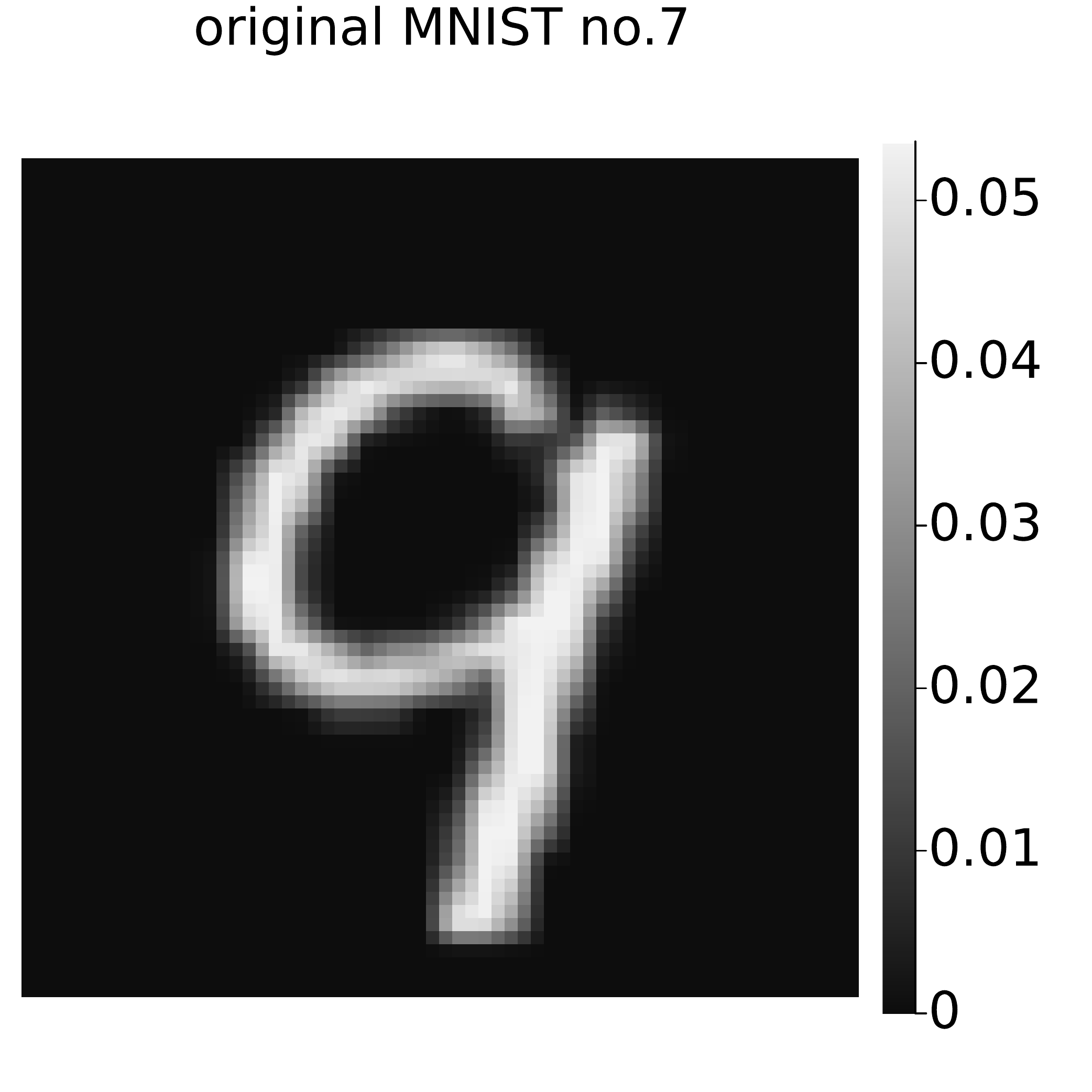}
    \end{minipage}
}
\subfloat[adversarial digit 9]{
    \begin{minipage}[h]{0.49\linewidth}
    \centering
    \includegraphics[width=1\linewidth]{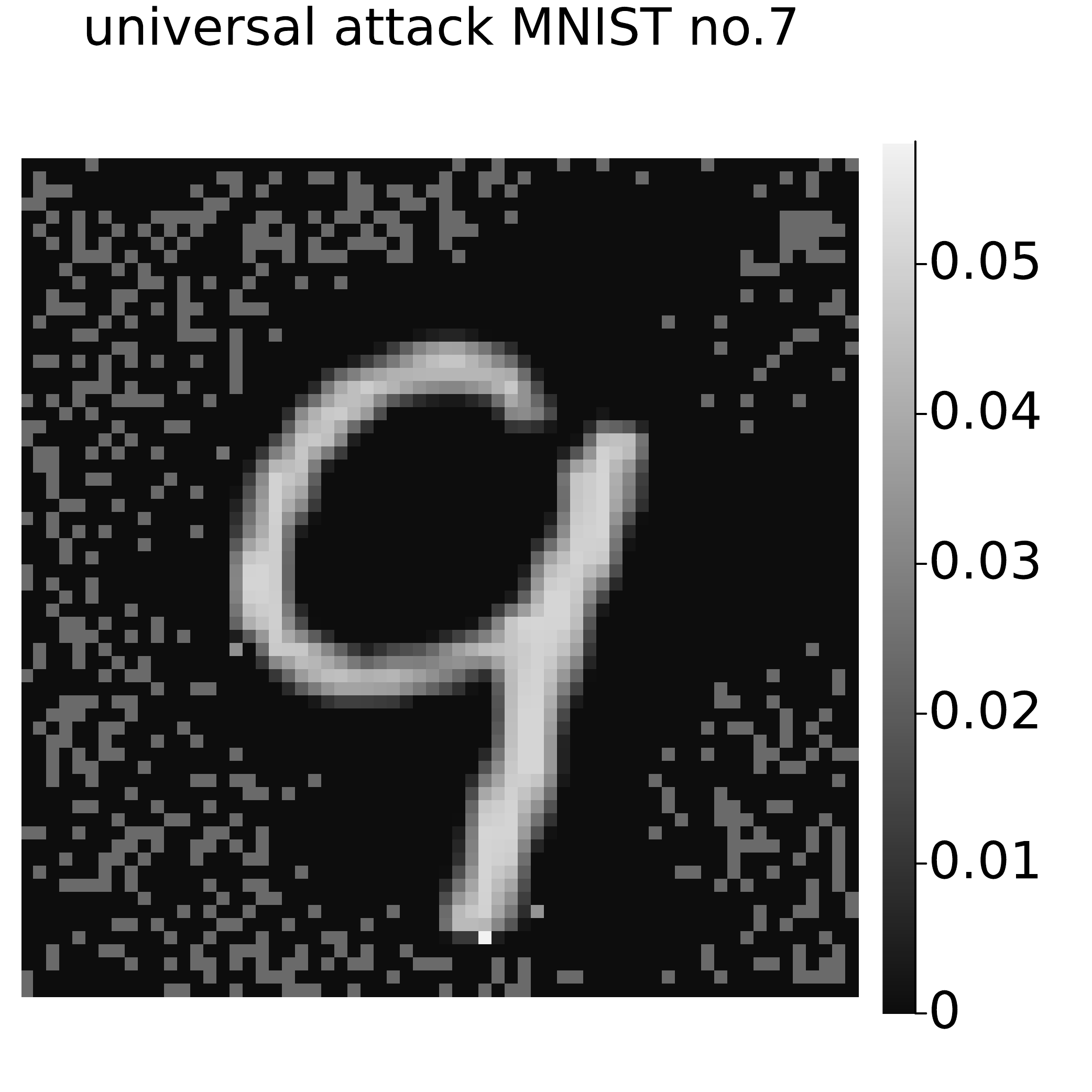}
    \end{minipage}
    
}
\qquad
\subfloat[MRI hand]{
    \begin{minipage}[h]{0.49\linewidth}
    \centering
    \includegraphics[width=1\linewidth]{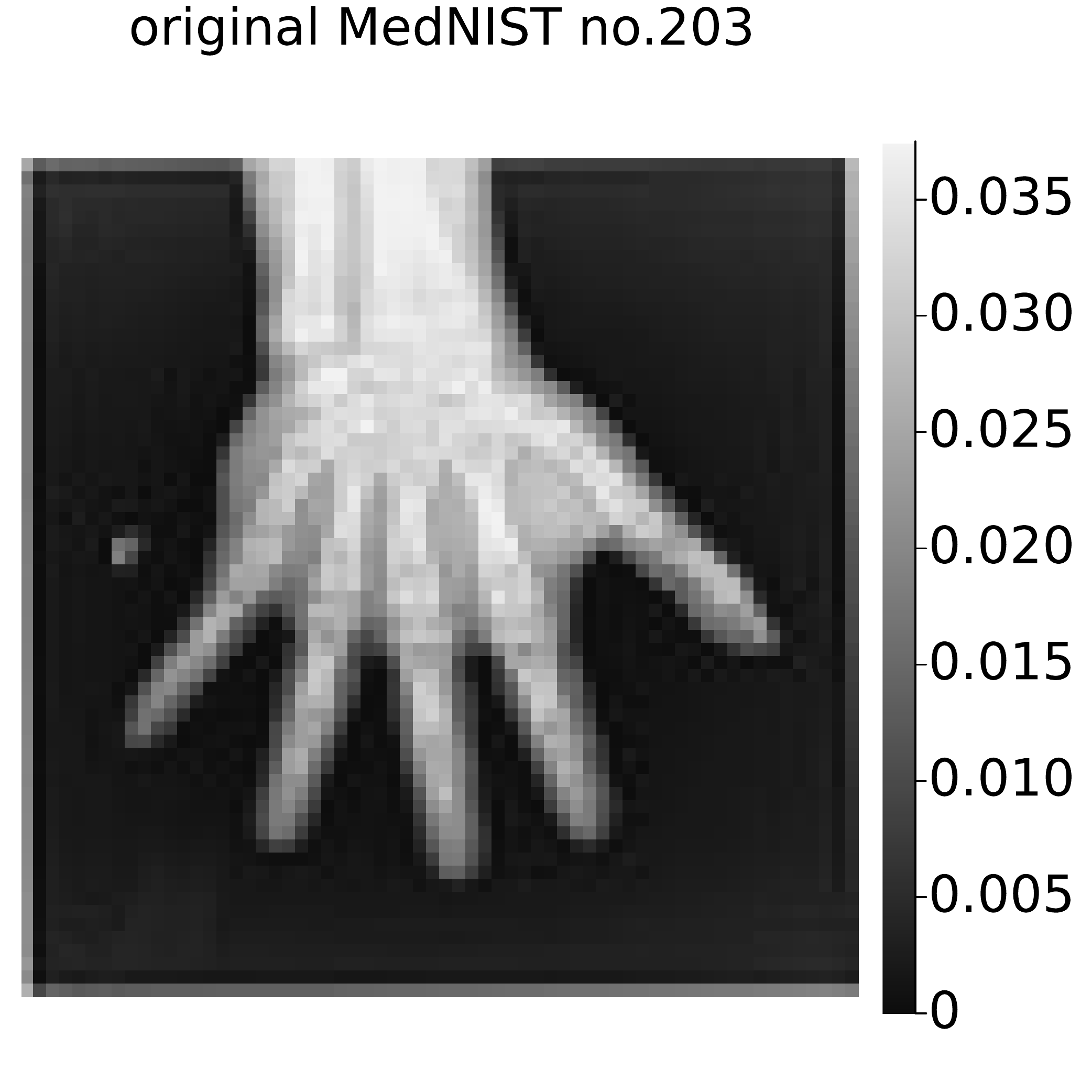}
    \end{minipage}
}
\subfloat[adversarial MRI hand]{
    \begin{minipage}[h]{0.49\linewidth}
    \centering
    \includegraphics[width=1\linewidth]{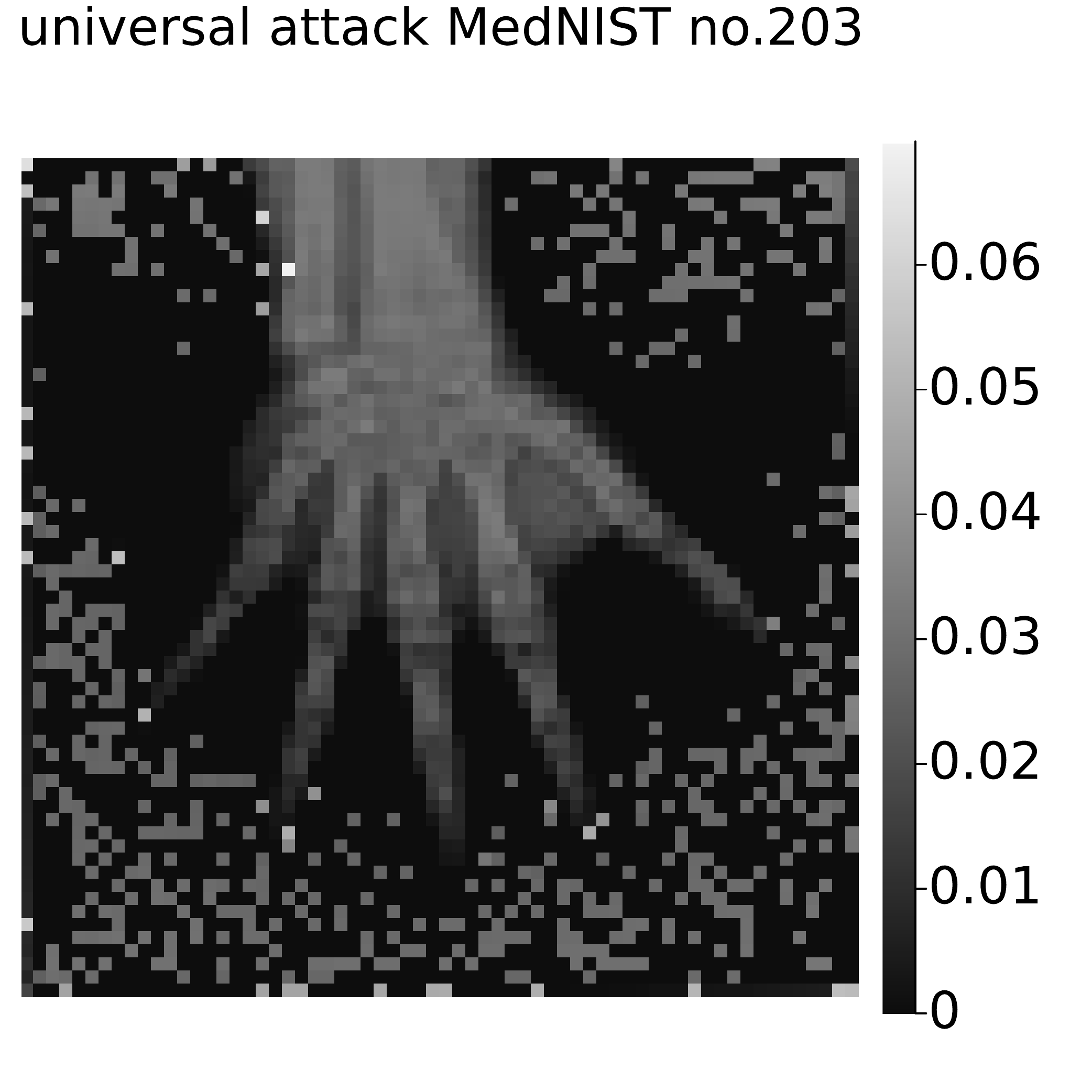}
    \end{minipage}
}
\qquad
\subfloat[MRI breast]{
    \begin{minipage}[h]{0.49\linewidth}
    \centering
    \includegraphics[width=1\linewidth]{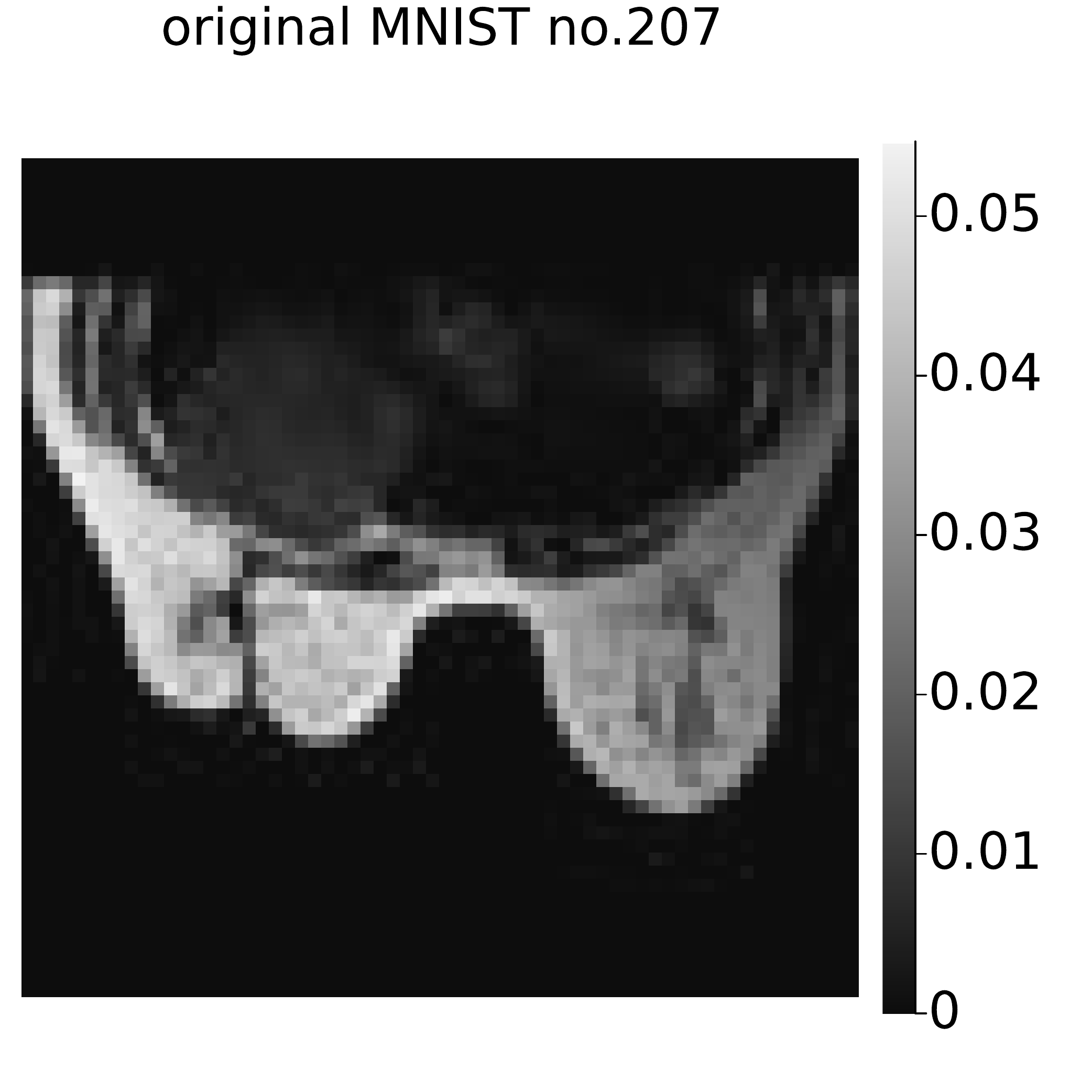}
    \end{minipage}
}
\subfloat[adversarial MRI breast]{
    \begin{minipage}[h]{0.49\linewidth}
    \centering
    \includegraphics[width=1\linewidth]{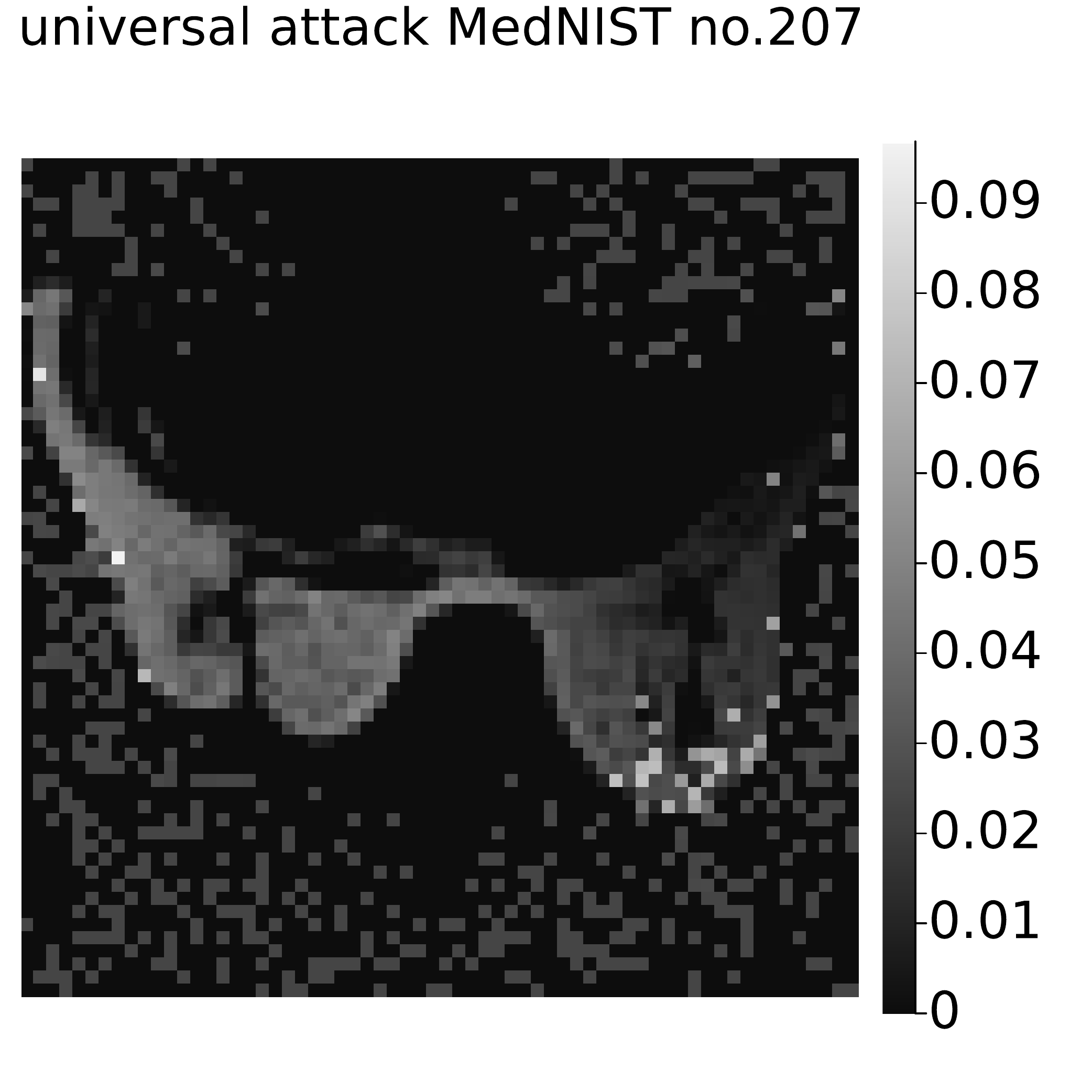}
    \end{minipage}
}
\captionsetup{justification=RaggedRight,singlelinecheck=false,textfont=normalsize,format=plain,indention=0pt}
\hyphenpenalty=-10
\tolerance=1000

    \caption{\textbf{Universal adversarial sample} This figure demonstrates some examples of quantum universal adversarial samples that the classifier wrongly classifies. Here we add the same perturbation to data samples from two datasets. Because of the requirements for a normalized state for the quantum classifier, the perturbation might be of a different shade in greyscale images to vision.}
        \label{fig:egfig3}
\end{figure}

Here, we plot some examples of universal adversarial samples after calculation in Fig. \ref{fig:egfig3}. For samples in MNIST handwritten digits, the distribution of non-zero value pixels is rather concentrated to the track of digit, thus the probability distribution after perturbing is almost unchanged except for the blurring at margin areas. However, the samples in the MedNIST dataset tend to distribute more evenly, resulting in a shallow color due to normalization steps. All of the adversarial samples listed above are wrongly classified into the other class by the quantum classifier with fidelity of around $51\%$ to $52\%$ which is just enough to deceive the classifier. We retain such a level of deceiving fidelity to minimize the perturbation and try to keep the fidelity value as high as possible.

{\color{black}
\section{Further Experiments}
The quantum classifier mentioned in previous paragraphs is the fully connected quantum classifier in particular. Meanwhile, all datasets used in numerical experiments are classical datasets previously. Such limitations might weaken our conclusion. Thus, to verify the effectiveness of the adversarial machine learning method in the above chapters, we also conducted further experiments on the quantum convolutional neural network(QCNN) classifier and the quantum data respectively.

\subsection{The QCNN scenario}
The QCNN classifier is a powerful classification model in the realm of quantum classifiers. According to the numerical experiment presented in Ref.\cite{oh2020tutorial}, the QCNN classifier is applied to $28\times28$ MNIST classification task. The QCNN classifier gets a similar performance to the classical CNN classifier and outperforms the fully connected classifier. This provides sufficient evidence that the QCNN classifier structure mentioned in Ref.\cite{oh2020tutorial} is a promising tool for classification tasks.

In our experiment, we follow Ref.\cite{oh2020tutorial} and construct a QCNN classifier with a similar structure on $12$ qubits. Specifically, our QCNN classifier contains $2$ convolutional layers, $2$ pooling layers, and $20$ fully connected layers. The convolutional layers find hidden states by applying rotational gates and controlled gates on adjacent qubits. The pooling layers use controlled rotational gates to reduce the qubits firstly from $12$ to $6$, then from $6$ to $3$. At last, the fully connected layers are applied with a depth of $20$ on the $3$ remaining qubits. During the decoding process, a measurement gate is applied to the last qubit of the circuit.

In the first training process, we set the learning rate to $0.005$ with $30$ training epochs. The dataset used for training is MNIST handwritten digits, the same as the previously mentioned experiment. A small fluctuation can be observed around epoch $10$, but the QCNN classifier reached a state-of-the-art accuracy above $95\%$ around epoch $15$. As mentioned in plenty of previous works, the QCNN classifier performs very well on a single classification task. Given the same datasets, hyper-parameter settings, and computing resources, the QCNN classifier trains much more quickly than the fully connected quantum classifier. Also, the QCNN classifier gets a lower loss than the fully connected quantum classifier after training, which leads to higher confidence in binary classification.

In the second training process, we keep the learning rate fixed at $0.005$, while changing the training steps to only $10$ training epochs to avoid over-fitting. Also, we raise the additional parameter $\lambda$ in the EWC method to $2000$ to protect the previously trained task. The training dataset is still the MedNIST hand and breast MRI images. After training, the QCNN classifier reaches an average accuracy of $91\%$ on both training tasks, which is $92.1\% $on the first classification task and $89.7\%$ on the second classification task. The training curve of both training processes of the QCNN classifier is presented in Fig.\ref{CNN_fig1}.

Here, we want to point out the potential threat of the QCNN classifier when applying it to continuous learning scenarios. Unlike fully connected quantum classifiers, the QCNN classifier contains a relatively small number of parameters. Intuitively, fewer training parameters lead to lower generalization capability of the given classifier. This intuition is perfectly verified in our experiment as we find out that the QCNN classifier can perform well on one training task but cannot maintain similar performance on multiple training tasks. Specifically, continuous drastic fluctuations can be observed in the training curves. We attempt to solve this issue by increasing the depth of the fully connected layer. Such an approach indeed alleviates the problem of insufficient training parameters but still requires careful adjustment of hyper-parameters.

% This issue is mainly generated by the limited computing ability of current quantum system simulations. For example, our experiment is conducted on an RTX-3060 GPU. Within a tolerable time span, 12 qubits are almost the maximal system size that we can perform. We attempt to solve this issue by increasing the depth of the fully connected layer. Such an approach indeed alleviates the problem of insufficient training parameters but still requires careful adjustment of hyper-parameters.

During the attacking process, we take similar hyper-parameter settings to the previous experiment, namely $0.02$ total perturbation strength, and $30$ perturbing iterations. The average accuracy of the two datasets decreased from $91\%$ to $34\%$ with fidelity of $81.5\%$. The accuracy of individual tasks dropped from 92.1\% and $89.7\%$ to $31.5\%$ (fidelity=$0.85$) and $36.5\%$ (fidelity=$0.78$). The attack process is drawn as a curve in Fig.\ref{CNN_fig2}. Such an attacking process of the QCNN classifier is consistent with the situation of the fully connected quantum classifier. This result indicates that the vulnerability of the QCNN and the fully connected quantum classifier is similar.

\begin{figure}[]
	\center
        \captionsetup{labelfont=large,textfont=large}
	\subfloat[The training process for the first task (QCNN)]{
        \begin{minipage}[t]{0.48\textwidth}
        \centering  
        \includegraphics[width=\linewidth]{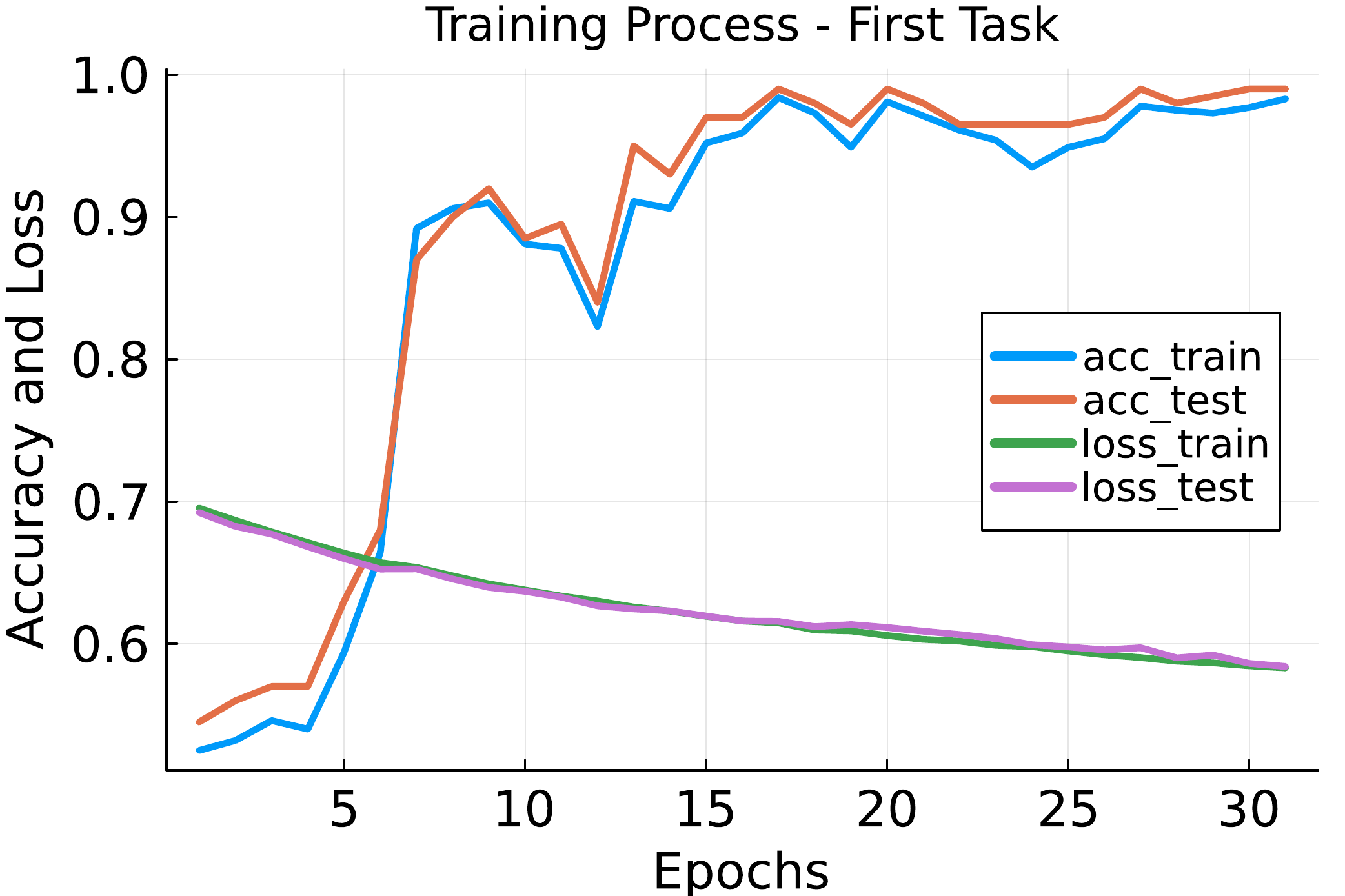}
        \end{minipage}
    }
    \qquad
    \subfloat[The training process for the second task (QCNN)]{
        \begin{minipage}[t]{0.48\textwidth}
        \centering  
        \includegraphics[width=\linewidth]{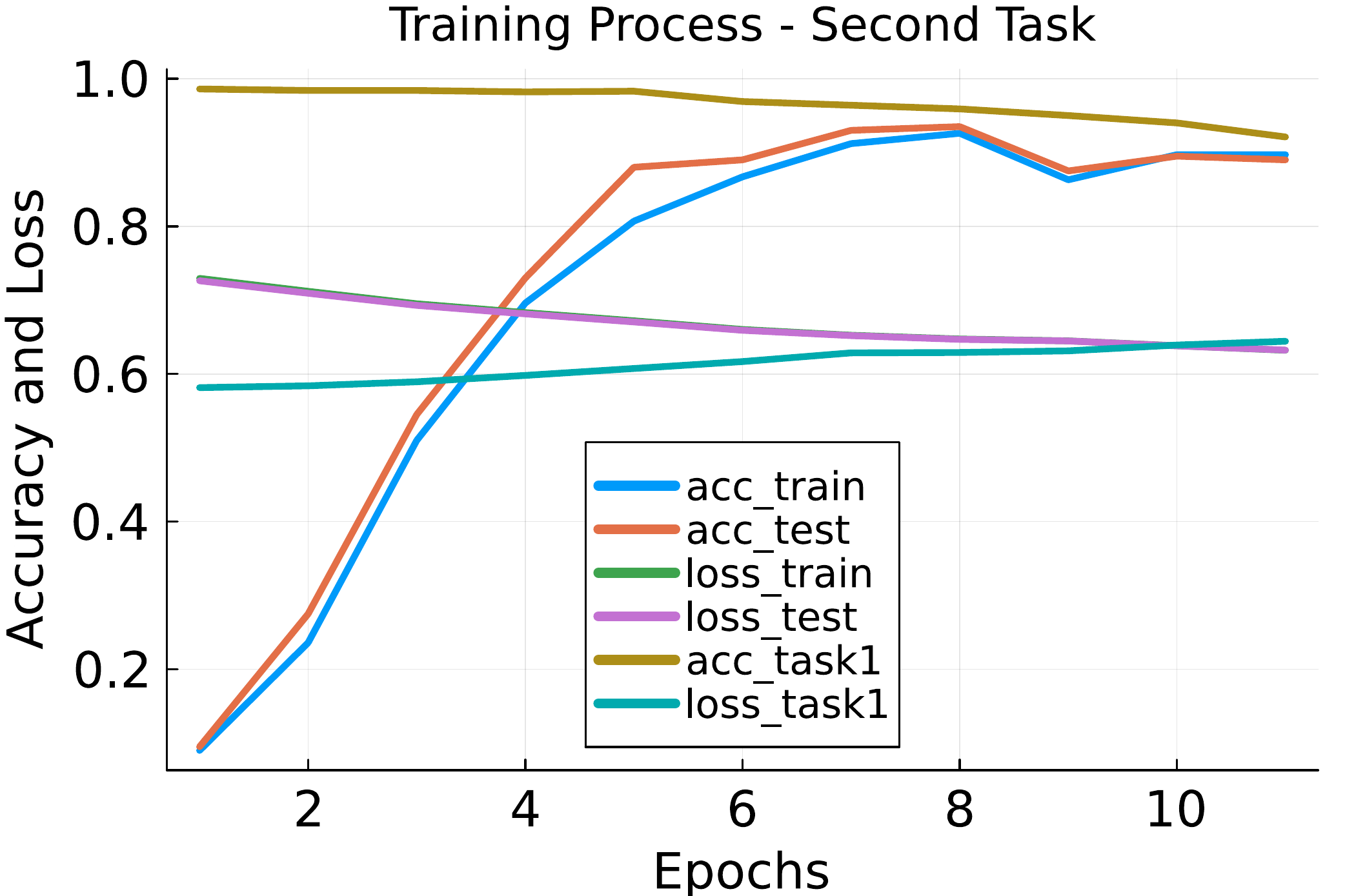}
        \end{minipage}
    }
\captionsetup{justification=RaggedRight,singlelinecheck=false,textfont=normalsize}
\vspace{-10pt}
    \caption{\color{black}
    \textbf{The training processes of quantum continual learning on the QCNN classifier.} (a) The first task is the classification of the MNIST hand-written digits which end up with an accuracy of around $95\%$. (b) The second task is the classification of the MedNIST MRI image training based on the optimized parameters of the first task. By utilizing the EWC method, we acquire an average accuracy of around $90\%$.}
 \label{CNN_fig1}
\end{figure}

\begin{figure}
   \centering
   \includegraphics[width=\linewidth]{
        {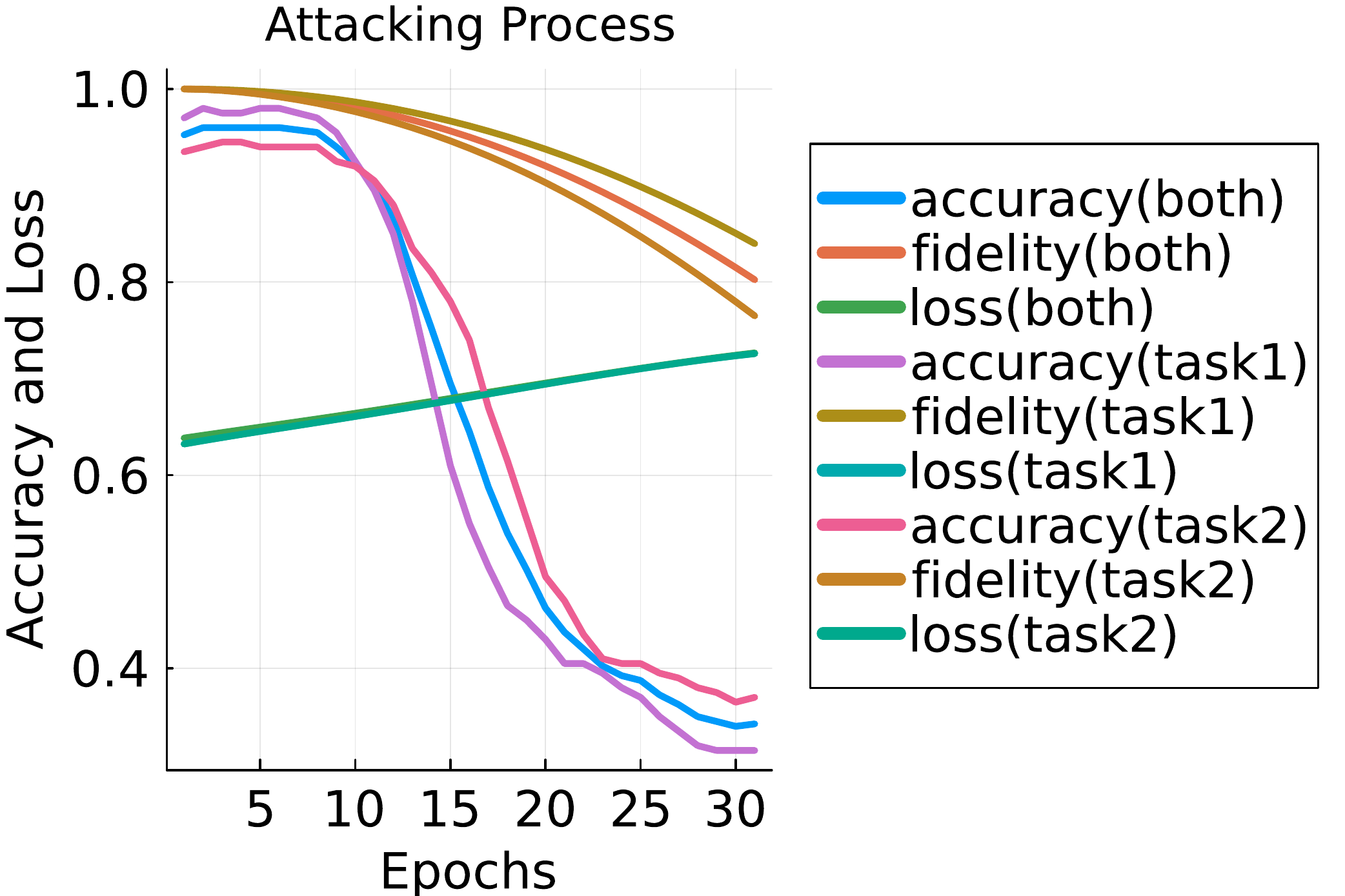}
    }
    \hyphenpenalty=-1
\tolerance=1000
\captionsetup{justification=RaggedRight,singlelinecheck=false,textfont=normalsize,width=\linewidth}
    \caption{\color{black}
    \textbf{Quantum universal perturbation on the QCNN classifier} This figure demonstrates the perturbing process of finding quantum universal perturbation on a QCNN classifier trained on the MNIST and the MedNIST datasets using quantum continual learning and EWC techniques. The average and separate accuracy, fidelity, and loss values are presented explicitly. There is not much difference in fidelity and loss, and accuracy value between different tasks.}
    \label{CNN_fig2}
\end{figure}

\subsection{Quantum data scenario}
The quantum classifiers are specially gifted to process quantum data. According to the evidence provided in Ref.\cite{jiang2019adversarial,li2022quantum}, quantum data can be efficiently learned by quantum classifiers and attacked by adversarial machine learning algorithms.

In our experiment, we choose a 12-qubit symmetry-protected topological (SPT) state dataset similar to Ref.\cite{li2022quantum} to verify the effectiveness of universal perturbation. Specifically, in the dataset containing SPT (Symmetry Protected Topological) states, we examine a one-dimensional cluster-Ising model featuring periodic boundary conditions. Suppose $\hat{\sigma}_{x}^{i}$, $\hat{\sigma}_{y}^{i}$, $\hat{\sigma}_{z}^{i}$ are Pauli matrices and $\lambda$ represents the relative strength of adjacent neighbour interaction. The Hamiltonian of this model can be put as the equation below:\cite{li2022quantum}
\begin{equation}
    H(\lambda)=-\sum_{j=1}^{N} \hat{\sigma}_{x}^{(j-1)} \hat{\sigma}_{z}^{(j)} \hat{\sigma}_{x}^{(j+1)} + \lambda\sum_{j=1}^{N} \hat{\sigma}_{y}^{(j)} \hat{\sigma}_{y}^{(j+1)}
\end{equation}

% I am not very sure about this paragraph

The model experiences a persistent quantum phase shift at $\lambda=1$, effectively distinguishing between two distinct phases. For $\lambda$ values less than $1$, it enters a cluster phase characterized by nonlocal hidden order. In contrast, for $\lambda$ values greater than $1$, it transitions into an antiferromagnetic phase with well-defined long-range order and a significant staggered magnetization. To comprehensively explore this transition, we systematically sweep $\lambda$ across the range from $0$ to $2$, utilizing intervals of $0.001$. We collect the resulting ground states at each interval, establishing them as our dataset for both training and testing purposes. We introduce the quantum data SPT in the second training process.

In the first training process, we keep the learning rate at $0.005$ and train for $30$ epochs on the MNIST handwritten dataset. The training is conducted on a fully connected quantum classifier on $12$ qubits with a depth of $20$. The training process generates state-of-the-art accuracy above $95\%$ on the MNIST handwritten dataset.

In the second training process, we keep the learning rate at $0.005$ but lower the training epochs to $20$. The additional parameter $\lambda$ in the EWC method is adjusted to $500$. The training dataset is the previously generated SPT dataset. After training, the fully connected quantum classifier reaches an average accuracy of $98\%$ on both training tasks, which is $98.6\%$ on the first classification task and $97.5\%$ on the second training task. The training curve of both training processes is presented in Fig.\ref{Qdata_fig1}.

% not sure if fidelity is appropriate here
During the attacking process, we lower the total perturbation strength to $0.015$ and the perturbing iteration to $15$. The average accuracy of the two datasets decreased from $98\%$ to $20.3\%$ with fidelity of $0.63$. The accuracy of individual tasks dropped from $98.6\%$ and $97.5\%$ to $4\%$  and $36.5\%$. The attack process is drawn as a curve in Fig.\ref{Qdata_fig2}. After introducing quantum data, the continuous model becomes more fragile than the cases before introducing quantum data. It can be observed in Fig.\ref{Qdata_fig2} that the second classification task on the SPT quantum dataset encounters a sharp decrease in the first attacking iterations. 
%Moreover, comparing Fig.\ref{fig:egfig2} and Fig.\ref{Qdata_fig2}, even though we lower the total perturbation strength from 0.02 to 0.015, the fidelity and accuracy value in Fig.\ref{Qdata_fig2} still cannot hold.

\begin{figure}[]
	\center
        \captionsetup{labelfont=large,textfont=large}
	\subfloat[The training process for the first task (quantum data)]{
        \begin{minipage}[t]{0.48\textwidth}
        \centering  
        \includegraphics[width=\linewidth]{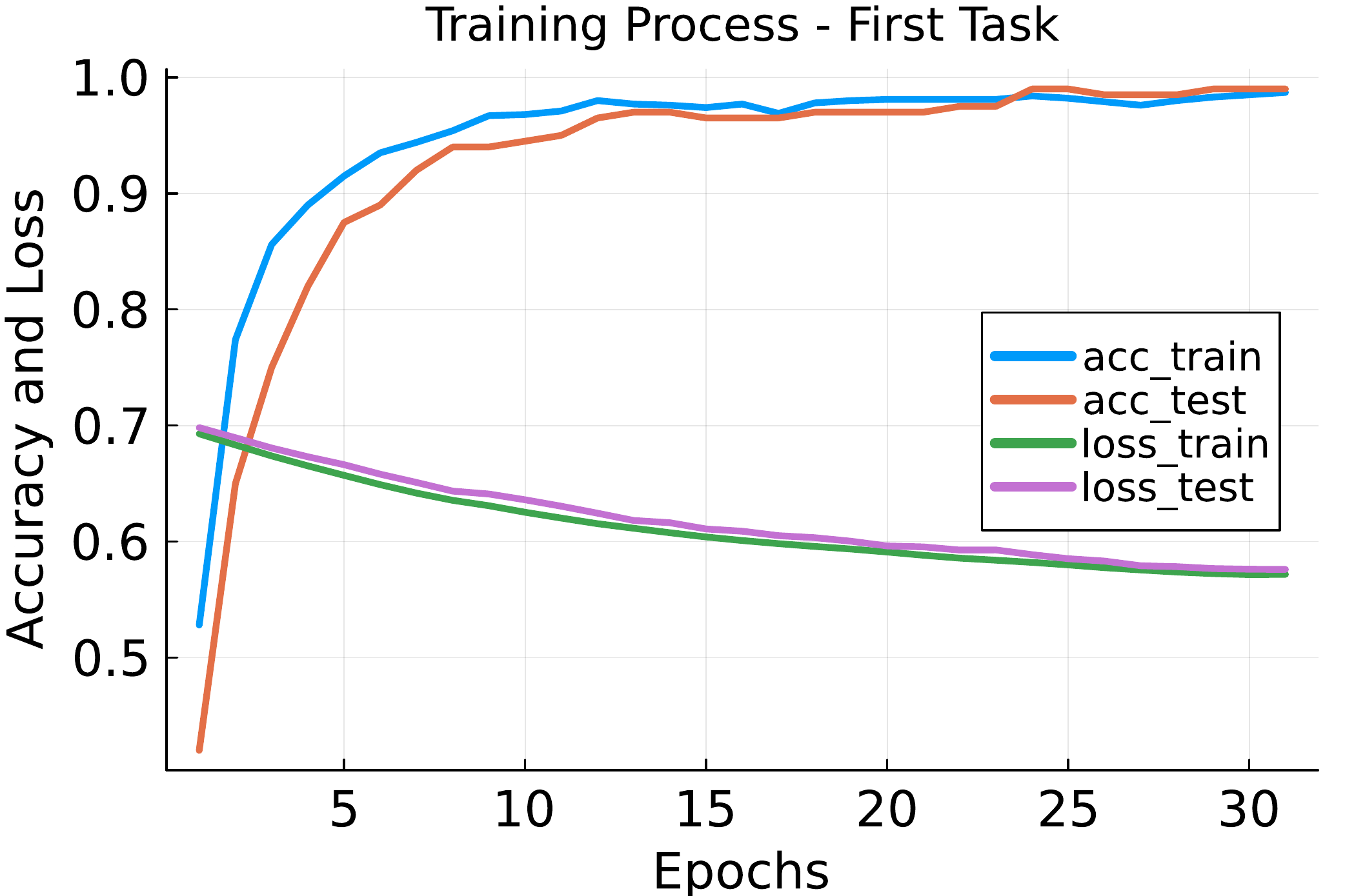}
        \end{minipage}
    }
    \qquad
    \subfloat[The training process for the second task (quantum data)]{
        \begin{minipage}[t]{0.48\textwidth}
        \centering  
        \includegraphics[width=\linewidth]{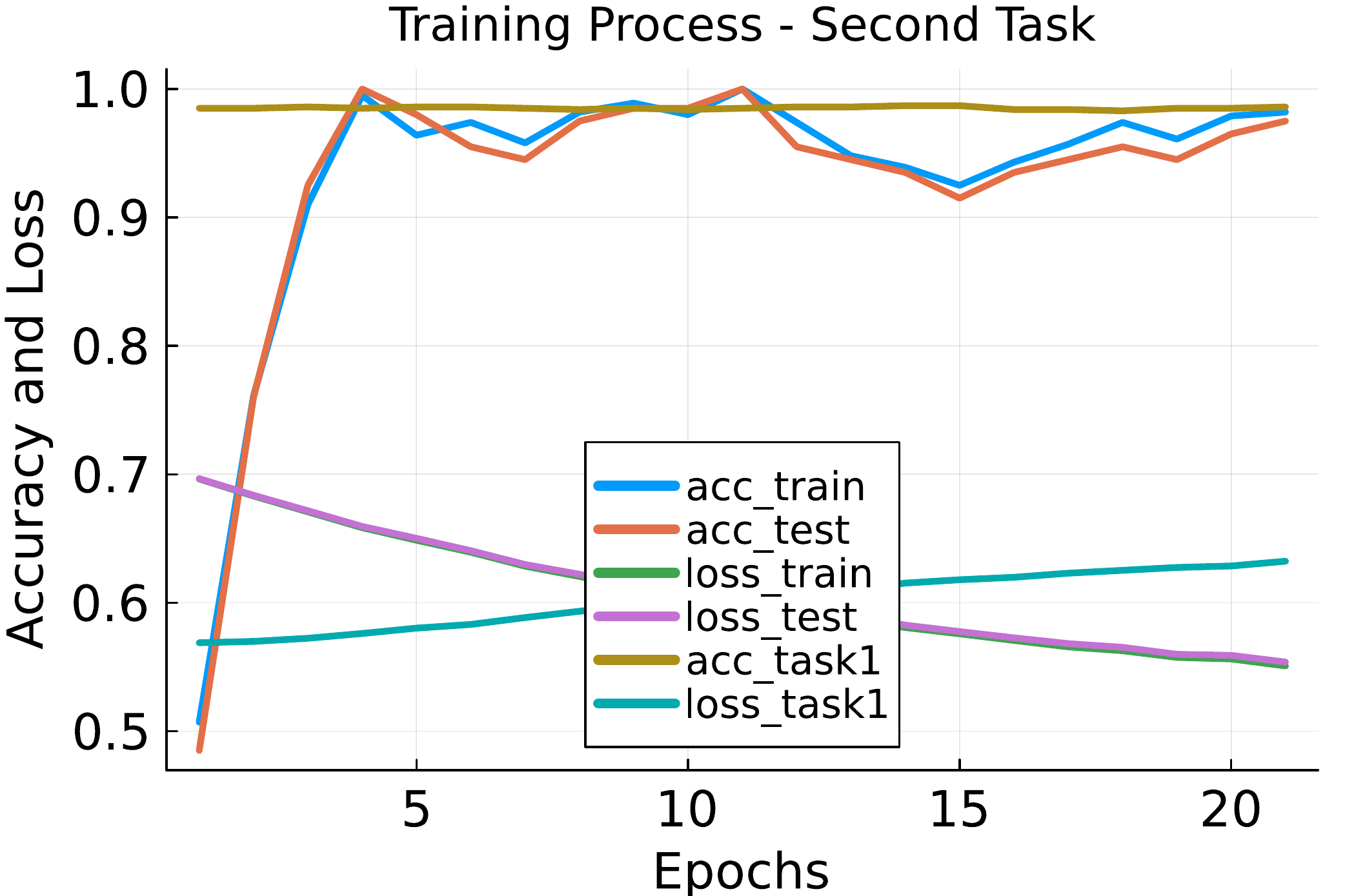}
        \end{minipage}
    }
\captionsetup{justification=RaggedRight,singlelinecheck=false,textfont=normalsize}
\vspace{-10pt}
    \caption{\color{black}
    \textbf{The training processes of quantum continual learning on quantum data.} (a) The first task is the classification of the MNIST hand-written digits which end up with an accuracy of around 95\%. (b) The second task is the classification of the SPT quantum dataset training based on the optimized parameters of the first task. By utilizing the EWC method, we acquire an average accuracy of above 90\%.}
 \label{Qdata_fig1}
\end{figure}

\begin{figure}
   \centering
   \includegraphics[width=\linewidth]{
        {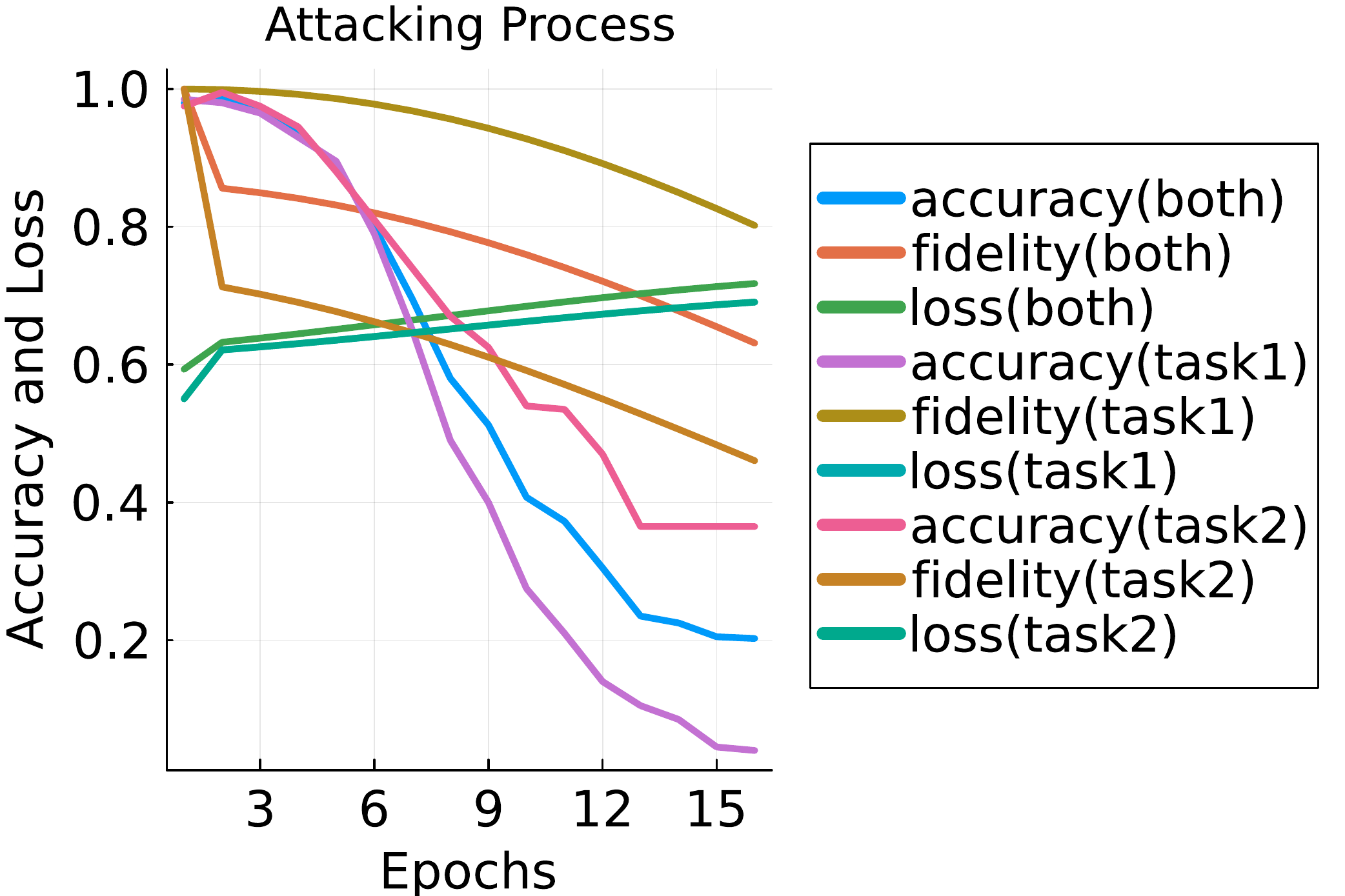}
    }
    \hyphenpenalty=-1
\tolerance=1000
\captionsetup{justification=RaggedRight,singlelinecheck=false,textfont=normalsize,width=\linewidth}
    \caption{\color{black} 
    \textbf{Quantum universal perturbation on quantum data} This figure demonstrates the perturbing process of finding quantum universal perturbation on a fully connected quantum classifier trained on the MNIST and the SPT datasets using quantum continual learning and EWC techniques. The average and separate accuracy, fidelity, and loss values are presented explicitly. There is not much difference in loss, but a prominent distinction in accuracy value between different tasks.}
    \label{Qdata_fig2}
\end{figure}

}

\section{Conclusion and Discussion}
This work presents a small glimpse into the emerging field of quantum adversarial machine learning, especially in finding universal perturbations. Typically, we first trained a classifier that can achieve almost state-of-the-art accuracy on two classification tasks with quantum continual learning techniques and elastic weight consolidation method to avoid catastrophic forgetting. The datasets of classification tasks are taken in binary form and encoded into quantum states. By applying the aforementioned quantum-adapted basic iterative method, we generated a universal perturbation that can deceive the classifier on both tasks conclusively with limited perturbation strength and comparably high fidelity. However, there are still numerous questions that are worth further exploration.

One of the challenges in this work is identifying which segment of the perturbation is useful. While the gradient consists of balanced information from the two classifiers trained on different tasks, it is difficult to precisely identify the boundary of a given classification task in practice. Mathematically characterizing this border remains an open question in the field. Based on previous numerical experiments in Ref. \cite{gong2022universal}, gradient information of one particular classification task can be generated into universal perturbation through likewise approaches. In this work, we found that the universal perturbation generated by gradient information of two different and independently distributed classification tasks can also deceive the quantum classifier successfully. It is essential to develop robust tools and techniques that can effectively discern the useful part of the universal perturbation generated by different tasks' gradients. Further research works addressing this issue might provide a clearer understanding of why universal perturbation can lower the accuracy of quantum classifiers.

Furthermore, it is important to note that the universal perturbation used in this work focuses specifically on supervised learning binary classification scenarios. This topic can be extended to multi-classification tasks and unsupervised or reinforcement learning scenarios. Developing effective methods for generating universal perturbations in these contexts is an ongoing research area. Currently, the existing study of quantum adversarial machine learning along this direction is relatively limited and requires further investigations. %There is a need for further exploration and development of techniques that can address the unique challenges posed by these learning paradigms and facilitate the application of universal adversarial attacks.

As observed from the above results, there were notable differences between the previously trained task and the later trained task during the perturbation process. The success of a quantum continual learning process lies in the application of the EWC method, which aims to protect crucial parameters to maintain the robustness of quantum classifiers across multiple training tasks. This suggests that certain circuit parameters hold greater importance compared to others. The previously trained classification task, having experienced a loss of specific important parameters, remains more vulnerable compared to the later trained classification task. This phenomenon might hint at a potential correlation between circuit parameters and classifier performance. Investigating the characteristics of continual learning in this context would be an intriguing avenue for further exploration.

It is worth noting that our work primarily focuses on classifiers with identical structures and datasets of equal sizes. However, in future endeavors involving universal quantum attacks, it would be valuable to explore attacking techniques that can be applied to classifiers with diverse structures and datasets of varying sizes. 

{\color{black}
The topic of defense strategy is also an interesting direction to be discovered. As far as we are concerned, adversarial training, randomized encoding, and quantum noise are several defense strategies that have proven to be useful for quantum adversarial attacks. Adversarial training defends adversarial attacks by putting adversarial samples into the training set so that the quantum classifier will be immune to similar attacks after training. In Ref.\cite{lu2020quantum}, the authors provide nice explanations and numerical experiments about this approach. Randomized encoding was also carefully discussed in Ref.\cite{gong2022enhancing}. By randomly encoding the legitimate data samples through unitary or quantum error correction encoders, the authors provide an interesting approach to protect quantum classifiers from adversarial attacks. Moreover, in Ref.\cite{du2021quantum}, authors also show that quantum noises can protect quantum classifiers from adversaries. However, adopting similar defense strategies in continual learning settings still needs substantial effort. Hence, we shall leave the discussion of defense strategy for future investigation.
}

%In summary, the preceding paragraphs demonstrate the existence of universal perturbation across different classification tasks. Also, we provide a straightforward approach to construct such universal perturbations through quantum continual learning and quantum basic iteration method. Nonetheless, there are still numerous unanswered questions listed in previous paragraphs that require further investigation so that we can deepen our understanding of quantum adversarial machine learning and universal perturbation.

\section{Acknowledgements}
We thank Dong-Ling Deng, Weikang Li, and Si Jiang for helpful inspiration and discussion. We acknowledge Weikang Li, Si Jiang, Wenjie Jiang, Weiyuan Gong, and Sirui Lu for discussing and sharing their numerical simulation codes. 

% \begin{appendices}
\appendix

\section{Specific Numerical Experiment Details}

\begin{figure*}
   \centering
   \includegraphics[width=0.4\linewidth]{
        {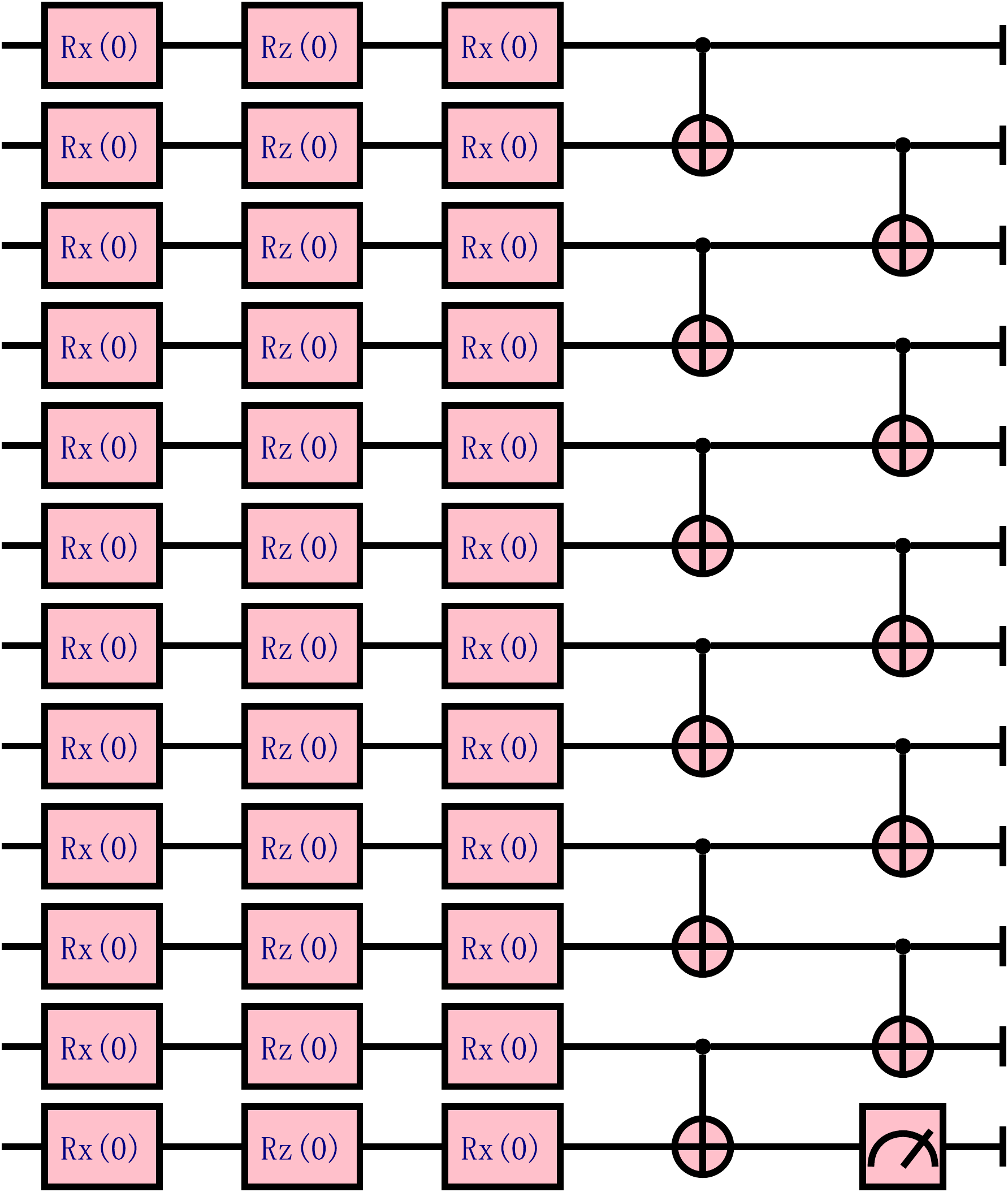}
    }
    \captionsetup{justification=RaggedRight,singlelinecheck=false,textfont=normalsize,width=\linewidth}
    \caption{\textbf{Quantum classifier layer illustration} This figure demonstrates one layer in the classifier mentioned above in the main text. Here we use Rx-Rz-Rx sequential gate structure as ``parameterized layer" and single layered controlled not gates for the ``entangled layer". The combination of such two layers is the unit of depth of the quantum circuit. The total circuit consists of $12$ qubits with a depth of $20$, while the measurement is taken on the last qubit.}
    \label{circuitdemo}
\end{figure*}

The quantum classifier used in this research was initially proposed by Ref. \cite{li2022quantum}. All numerical results are performed on Yao.jl framework \cite{luo2020yao,bezanson2012julia}.  

Yao is a highly efficient computing software that simulates quantum computing behavior with automatic differentiation and provides a vast range of powerful assistant interfaces for research. Now, such a tool is attracting more and more researchers around the globe.

Typically, for classical image data like MNIST and MedNIST, we need to encode them into logical quantum states before putting them into the quantum circuit for calculation. We are using amplitude encoding for the $64\times64$ rescaled input data because this encoding technique can greatly reduce the quantum qubit space complexity from $\Theta(N)$ to $\Theta(\mathrm{log}(N))$. Considering the lack of computational resources of qubit size, reducing the number of necessary qubits is a very important philosophy to enhance the potential computational power of a quantum circuit. To implement amplitude encoding, we simply create a ``zero-state" quantum state of size 12 and assign the initial value in matrix form through the ``.state" interface.

An illustration of a section of our trained quantum circuit is shown in Fig.\ref{circuitdemo}. Specifically, we put one ``Parameterized Layer" and one ``Entangled Layer" together as one depth of our quantum circuit. In the ``Parameterized Layer", we apply sequential quantum rotation gates on X-Z-X respectively. For arbitrary parameter $\theta$, gate $R_X$ is equivalent to $e^{i\theta/2X}$, and gate $R_Z$ is equivalent to $e^{i\theta/2Z}$. For every gate in the quantum circuit, we take the angles $\theta$ as the parameters to be optimized in the training process. Such structures were reported and tested to be robust and hardware-efficient in Ref. \cite{li2022quantum}. The ``Entangled Layer" on the other hand is a composite of single-layered controlled not gates to create entanglement between different qubits. As mentioned in Ref. \cite{ugwuishiwu2020overview,bulger2003implementing}, entanglements are a vital key to the great potential of quantum computing.

Here, we take measurements on the last qubit of our quantum circuit. Without a doubt, similar measurements can also be made on other qubit indexes with similar philosophies. Speaking of measurement, we want to highlight that we are using a threshold of probability of $0.5$. In other words, for binary implementation, we set two measurement operators, namely $op0$ and $op1$. The classification decision is made by determining on which operator the measured probability exceeds the threshold level. More interestingly, we provide insight into performing multiple classifications in this context by setting multiple measuring operators and multiple measuring qubits. Technically, provided that we are doing a $N$ classification task, we can take measurements on $\mathrm{log}(N)$ number of qubits and discern them in a similar way.

In order to achieve a stable performance, we take 12 qubits and 20 composite layers to provide comparably sufficient parameters for initial training and quantum continual training. Based on our numerical experiments and duplicate verification, such a choice of parameters can provide a robust and efficient quantum classifier.

{\color{black}
The QCNN classifier refers to a similar technical routine with fully connected quantum classifiers, the structure can be seen in Fig.\ref{QCNNstructure}.

\begin{figure*}
   \centering
   \includegraphics[width=\linewidth]{
        {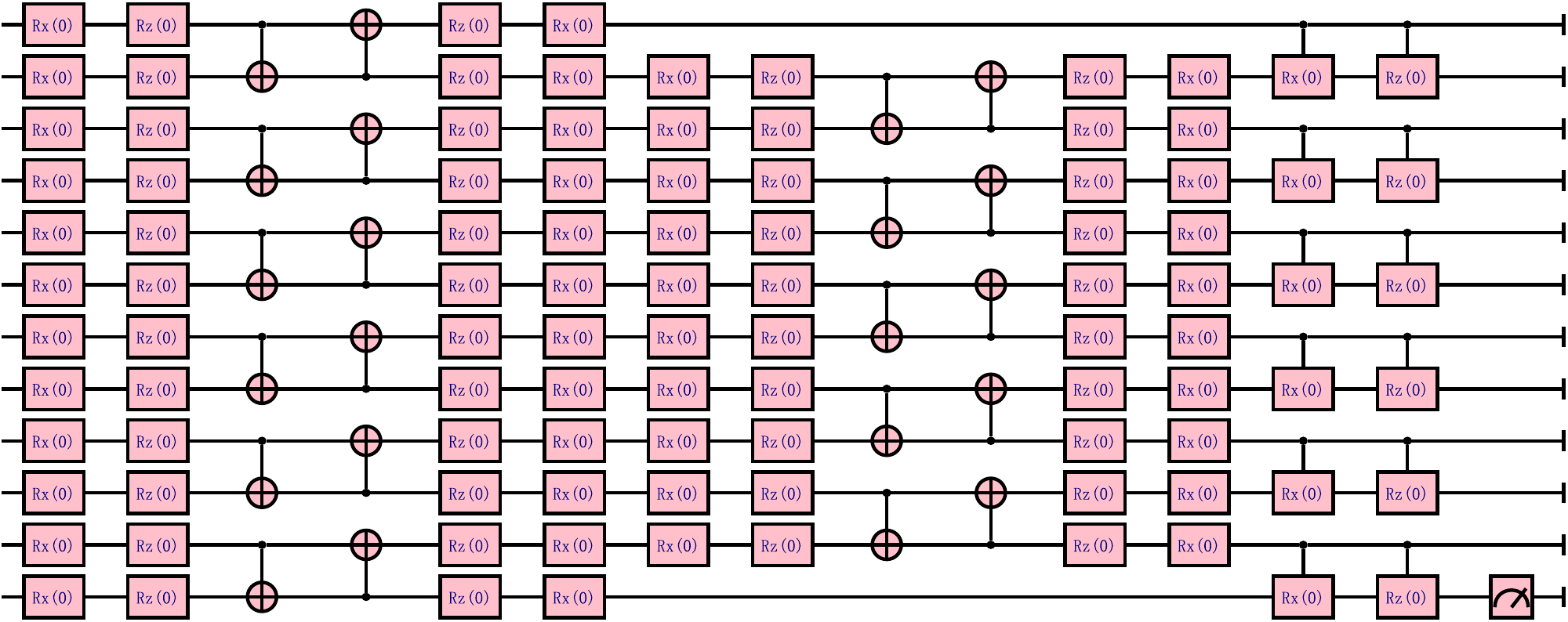}
    }
    \captionsetup{justification=RaggedRight,singlelinecheck=false,textfont=normalsize,width=\linewidth}
    \caption{\color{black}
    \textbf{QCNN classifier layer illustration} This figure demonstrates one convolutional and pooling layer in the QCNN classifier mentioned above in the main text. Here we apply rotational gates and controlled gates to adjacent qubits to perform convolutional calculation. After the pooling layer, the size of the working qubit shall be reduced to half of the original size.}
    \label{QCNNstructure}
\end{figure*}

}

\section{Technical details for EWC method}
  In the above sections, we provide a brief explanation of the elastic weight method. To be specific, we provide some technical details for the EWC method here in the appendix. Suppose the Gaussian distribution equation is defined as $N(\mu,\sigma)$ and we use a function $f(\theta)$ to mark the logarithmic value of this probability, then we can do the calculation as follow:
\begin{equation}
    P(X_A|\theta) = \frac{1}{\sqrt{2\pi}\sigma}e^{-\frac{(\theta - \mu)^2}{2\sigma^2}}
\end{equation}

\begin{equation}
     f(\theta) = \mathrm{log}P(X_A|\theta)
\end{equation}

Let the optimal solution be $\theta_A$, then the equation after Taylor expansion is as follows:

\begin{equation}
    \begin{cases}
        f'(\theta_A)=0 \\
        f(\theta) = f(\theta_A)+f'(\theta_A)(\theta-\theta_A)+f''(\theta_A)\frac{(\theta-\theta_A)^2}{2}
    \end{cases}
\end{equation}

Since the element $\mathrm{log}\frac{1}{\sqrt{2\pi}\sigma}$ and element $f(\theta_A)$ are constant, we can simplify the above equation as follow:

\begin{equation}
    -\frac{(\theta-\mu)^2}{2\sigma^2}=f''(\theta_A)\frac{(\theta-\theta_A)^2}{2}
\end{equation}
Then after the above calculation we can get results: $\mu=\theta_A$ and $\sigma^2 = -\frac{1}{f''(\theta_A)}$. Now we can simulate the posterior probability with an acceptable error range. According to Bayes's rules, the distribution form of $p(\theta|X_A)$ and $p(X_A|\theta)$ is the same, the proof is trivial. But this form is not efficient to calculate, more transformations need to be applied.

The Fisher information matrix \cite{vsafranek2018simple,spall2005monte,ly2017tutorial} is the covariance of the probability distribution gradient. Officially, the Fisher information is a measurement of the amount of information that an observable random variable $X$ carries about an unknown parameter $\theta$ of a distribution that models $X$. For batched input $X={X_1,X_2,\ ...\ ,X_n }$, the Fisher information matrix can be taken down as:
\begin{equation}
    F=\frac{1}{n}\sum_{i=1}^n \nabla\mathrm{log}P(X_i|\theta)\nabla\mathrm{log}P(X_i|\theta)^T
\end{equation}

Unfortunately, such a form is still not friendly for calculation, we can further simplify the above equation using the Hessian matrix. The Hessian matrix \cite{thacker1989role,mizutani2008tutorial} $H$ is a square matrix that is a composite of second-order partial derivatives of a scalar-valued function, describing the local curvature of a function of many variables. For machine learning, the Hessian matrix is very easy to acquire, the question is how to link the Hessian matrix and the Fisher information matrix. The answer is that the Fisher information matrix is the negative expectation of the Hessian matrix. To prove this, we can take two steps: transform the Hessian matrix $H$ and calculate its expectation. The proof is as follows.

First, we rewrite the Hessian matrix in a novel form of partial derivatives:

\begin{eqnarray}
    \begin{aligned}
        & H_{\mathrm{log}P(X|\theta)} = J(\nabla\mathrm{log}P(X|\theta)) 
        =J(\frac{\nabla P(X|\theta)}{P(X|\theta)}) \\
        &=\frac{H_{P(X|\theta)}P(X|\theta) - \nabla P(X|\theta)\nabla P(X|\theta)^T}{P(X|\theta)^2} \\
        &=\frac{H_{P(X|\theta)}}{P(X|\theta)} - (\frac{\nabla P(X|\theta)}{P(X|\theta)})(\frac{\nabla P(X|\theta)}{P(X|\theta)})^T
    \end{aligned}
\end{eqnarray}
Second, we calculate the expectation of the Hessian matrix in the partial derivative form:

\begin{eqnarray}
    \begin{aligned}
        & \underset{P(X|\theta)}{\textbf{E}}[\frac{H_{P(X|\theta)}}{P(X|\theta)}-(\frac{\nabla P(X|\theta)}{P(X|\theta)})(\frac{\nabla P(X|\theta)}{P(X|\theta)})^T] \\
        &=\int\frac{H_{P(X|\theta)}}{P(X|\theta)}P(X|\theta)d\theta - 
        \underset{P(X|\theta)}{\textbf{E}}[\nabla\mathrm{log}P(X|\theta) \nabla\mathrm{log}P(X|\theta)^T] \\
        & = \int H_{P(X|\theta)}d\theta-F = -F
    \end{aligned}
\end{eqnarray}

By applying an intuitive mathematical trick, we successfully find a solution to link the Hessian matrix and the Fisher information matrix.

On the other hand, for 1-dimensional functions, the Hessian matrix has property as follows:
\begin{equation}
    H_{\mathrm{log}P(X|\theta)} = (\mathrm{log}P(X|\theta))''
\end{equation}

Thus, for the Fisher information matrix of parameter $\theta_i$ and dataset with $n$ samples, we can rewrite in second derivative form as follow:
\begin{equation}
    \begin{aligned}
        F_i = -\underset{P(X|\theta)}{\textbf{E}}[H_{\mathrm{log}P(X|\theta)}] = -\frac{1}{n}\sum f_i''(\theta_{A,i})
    \end{aligned}
\end{equation}

For easier calculation, we import fisher information matrix $\mathbf{F}$ and hyper-parameter $\lambda$ to evaluate the importance of element2 to element1. We can put the calculation above as:
\begin{equation}
\begin{aligned}
     &\underset{\theta}{\mathrm{min}}\ (\mathrm{log}p(\theta|X_A)) = \underset{\theta}{\mathrm{min}}\ (f''(\theta_A)\frac{(\theta-\theta_A)^2}{2}) \\
    &= \underset{\theta}{\mathrm{min}}\ (\frac{\lambda}{2}\sum_i \mathbf(F)_i(\frac{(\theta{i}-\theta_{Ai})^2}{2}))
\end{aligned}
\end{equation}

% \end{appendices}

 % \vspace{10mm}

% \bibliographystyle{iopart-num.bst}
\bibliography{main.bib} 

\end{document}